\documentclass[a4paper,11pt]{article}

\usepackage{graphicx}
\usepackage{amsmath}
\usepackage{amsfonts}
\usepackage{amssymb}
\usepackage{simplewick}
\usepackage{a4wide}
\def\Dirac{{\raise0.09em\hbox{/}}\kern-0.69em D}% Dirac operator
\def\ep{i\epsilon} 

\def\lesssim{\mathrel{\hbox{\rlap{\hbox{\lower4pt\hbox{$\sim$}}}\hbox{$<$}}}}
\def\sq{\hbox{\rlap{$\sqcap$}$\sqcup$}}         % d'Alembertian
\def\p{\partial}                                %\partial derivative 
\def\tfrac #1#2{\textstyle{\frac{#1}{#2}}} 	% textstyle fraction
\def\dfrac #1#2{\displaystyle{\frac{#1}{#2}}} 	% displaystyle fraction
\def\tr{\mbox{Tr}\,}                            % trace 

\def\beg{\begin{eg}\rm}                         % Macro to begin an Example
\def\eeg{\hfill\sq\end{eg}}                     % Macro to end an Example

\def\cS{{\cal S}}

\def\k {\kern-.1em\mathbin{,}\kern-.1em}
\def\hk{\kern.12em\raise-1em\hbox{$\hat{\raise1em\hbox{,}}$}\kern.12em}

\def\exterior{{{\raise0.2em\hbox{$\scriptstyle\bigwedge$}}{}}}

\def\tF{{\sf F}}

\def\tX{{\sf X}} \def\tx{{\bar x}}

\def\hx{{\hat x}}
\def\hp{{\hat  p}}

\def\tK{{\tilde K}}

\def\ds{\stackrel{\star}{,}}
\def\cK{{\cal K}} 
 \def\cN{{\cal N}}
\def\cJ{{\mathbb{ J}}}

\def\bbN{{\mathbb{N}}}
\def\bbK{{\mathbb{K}}}
\def\bbKK{{\mathbb{K}\mathbb{K}}}
\def\bbQ{{\mathbb{Q}}}
\def\bbB{{\mathbb{B}}}

\newcommand{\initiate}{\setcounter{equation}{0}}        
\begin{document}

\title{Quantization of a gauge theory \\ on a curved noncommutative space}

\vskip30pt
\author{
M.~Buri\' c$^{1}$\thanks{majab@ipb.ac.rs},\  \ 
M.~Dimitrijevi\' c$^{1}$\thanks{dmarija@ipb.ac.rs}, 
V.~Radovanovi\' c$^{1}$\thanks{rvoja@ipb.ac.rs}\\
and M.~Wohlgenannt$^{2}$\thanks{michael.wohlgenannt@univie.ac.at}
  \\[25pt]
\\$\strut^{1}$
       University of Belgrade,       Faculty of Physics, \\ Studentski trg 12,
       SR-11001 Belgrade 
 \\[10pt]$\strut^{2}$
		   University of Vienna, Faculty of Physics\\ Boltzmanngasse 5, A-1090 Vienna}  
\maketitle

\begin{abstract}
We study quantization of a gauge analogon of the Grosse-Wulkenhaar model: we 
 find divergent one-loop contributions to 1-point and
2-point Green functions. We obtain that five counterterms are necessary for
renormalization and that all divergences are logarithmic.
\end{abstract}

\initiate
\section{Introduction}

Although the renormalization of quantum field theory was the main objective of
 Snyder \cite{Snyder:1946qz} and the  others at the time when 
 noncommutativity of coordinates was first introduced, its full
understanding  in the context of noncommutative quantum field theory 
and the search for renormalizable models are still open issues.
This applies in particular to theories defined on the  Moyal space.
When one  deforms a commutative field theory  to a noncommutative one
 by replacing the ordinary by the Moyal-Weyl product,
 a common pattern appears: in addition to the usual 
divergences, the new  model suffers from the so-called 
ultraviolet-infrared  (UV/IR) mixing. In the case  of scalar field the additional 
divergences can be handled by a modification of the propagator.  
Two different possibilities are known:  one can  add  either
a position-dependent  potential term $\, (\bar  x\star \phi)^2\,$
 \cite{Grosse:2003nw,Grosse:2004yu}, or a nonlocal kinetic term
$\, \phi\, {\Box}^{-1}\phi\, $  \cite{Gurau:2008vd} to the
original action. The resulting theories differ in many
respects but both  are renormalizable  to all orders in perturbation theory. 

We will follow and develop the first approach. The Grosse-Wulkenhaar (GW) 
model is defined by 
\begin{align}
\label{eq}
  \cS=\int  \frac{1}{2}\,\partial_\mu\phi\star\partial^\mu\phi+\frac{\mu_0^2}{2}\,\phi\star\phi
 +\frac{\Omega^2}{4}(\tx\phi)\star(\tx\phi)+\frac{\lambda}{4!}\, \phi\star\phi\star\phi\star\phi ,
\end{align}
where $\tx_\mu=\left(\theta_{\mu\nu}\right)^{-1}x^\nu$, 
$ \theta_{\mu\nu}$ is a constant noncommutativity  tensor,
 $\, [x^\mu\ds x^\nu] = i\theta^{\mu\nu}\, $, and the $\star$ is the Moyal-Weyl 
star product (\ref{Moy}). The model is fully renormalizable;  
in addition, a nontrivial fixed point occurs for $\Omega=1$ at which the 
$\beta$-function for the coupling constant vanishes.
One can understand renormalizability of  (\ref{eq}) physically 
either as a consequence of an additional symmetry called the 
Langmann-Szabo duality \cite{Langmann:2002cc} which the model has, 
or as a consequence of the  confinement around the origin of coordinates
 introduced through the quadratic potential. 

Despite numerous attempts to generalize the Grosse-Wulkenhaar model
 to gauge theories a similarly successful gauge model
has not been by now found. A difficult point 
roughly speaking is, how to include the oscillator or an
analogous position-dependent term in a gauge invariant way.
In a straightforward approach \cite{deGoursac:2007qi,Grosse:2007dm},
one couples the scalar field 
\eqref{eq}  to an external gauge field: the dynamics of the gauge field 
is then extracted from the divergent contributions to the one-loop effective action. 
The induced gauge field action however contains the explicit tadpole terms and  gives 
 rise to a non-trivial vacuum, \cite{deGoursac:2008rb}.
Another, simplified version of a gauge model was discussed in
\cite{Blaschke:2007vc,Blaschke:2009aw}.
The model includes only the oscillator potential for the gauge field while the tadpole
 terms are omitted. Hence the considered action is not gauge invariant:
 nonetheless the BRST invariance can be established by an appropriate
 choice of ghosts and auxilliary fields.
 Although  tadpoles are not present at the tree-level in quantization
 they reappear as the UV-counterterms at one loop. 

A different line of generalization of the GW model was proposed in
 \cite{Buric:2009ss,Buric:2010xs}.
 It is based on a geometric interpretation of action (\ref{eq})
as a  dimensionally reduced 
 action for a scalar field on curved noncommutative  space.
The oscillator potential is then not external but  it is rather 
 the coupling to the  curvature. This framework gives a natural prescription 
 to define the action for gauge fields with position-dependent couplings
while preserving the gauge symmetry. A specific
feature of the model is that in dimensional reduction one of the 
gauge degrees of freedom becomes a scalar field. The model has  two classical vacua
 one of which is trivial, $\phi=0$, $A_\alpha =0$, and suitable for quantization. 
Moreover, the BRST invariance of the gauge-fixed
action can be easily established.

We initiate here the study of perturbative quantization of
 the described model \cite{Buric:2010xs}.
The outline of the paper is the following.
In Section~2 we introduce our approach
that is we recollect all relevant concepts and formulae. 
In Section~\ref{sec:quantization} we derive the 
Feynman rules for the  propagators and the  vertices in  momentum space.
We obtain in Section~4 that the one-loop tadpole corrections do not vanish and
we calculate the corresponding divergent counterterms. In Section~5 we find the
one-loop divergent propagator corrections. In the concluding section we discuss our
 results and the work which remains to be done in the future.

\initiate
\section{The classical model}

Our gauge model is obtained by dimensional reduction from  three-dimensional noncommutative space 
called  the truncated Heisenberg algebra. The algebra is generated by  hermitian
coordinates, operators  $\,\hx^1$, $\hx^2$, $\hx^3$ which satisfy commutation relations 
\begin{eqnarray}
 &&  [\mu \hx^1,\mu \hx^2] =i\epsilon (1-\bar\mu \hx^3)  ,\label{trunc}  \\
&&       [\mu \hx^1,\bar\mu \hx^3] = i\epsilon (\mu \hx^2 \bar\mu \hx^3+\bar\mu \hx^3 \mu \hx^2)  ,   
 \nonumber      \\
&&[\mu \hx^2,\bar\mu \hx^3] = - i\epsilon (\mu \hx^1\bar\mu \hx^3+\bar\mu \hx^3 \mu \hx^1)  ,  \nonumber
\end{eqnarray} 
where $\epsilon$ is  dimensionless noncommutativity parameter while $\mu$ and $\bar\mu$ have 
dimension of mass. Unlike in \cite{Buric:2010xs}, here we will denote  the noncommuting
variables in a generic situation that is when  representation is not specified by a hat,
 keeping the corresponding unhatted characters  for the
corresponding quantities in the Moyal-space representation.
Algebra (\ref{trunc}) besides the usual commutative  limit $\epsilon\to 0$
 has  an interesting contraction $\bar\mu\to 0$ to the Heisenberg algebra. 
For $\epsilon =1$  (\ref{trunc}) has finite-dimensional $n\times n$ matrix
representations which in the limit  $\bar\mu\to 0$ (or equivalently $n\to\infty$)  tend to the known
unique infinite-dimensional representation of the Heisenberg algebra. We will refer to this limit as 
to   dimensional reduction from three-dimensional truncated Heinsenberg algebra to its two-dimensional
 subspace $\hx^3=0$; this interpretation, as  it was shown in \cite{Buric:2010xs},  is consistent not only
at the level of algebra  but also at the level of exterior algebra. In 
order to avoid two mass scales in the model we will also put  $\bar\mu =\mu$ and discuss only
this case.

The truncated Heisenberg algebra is a smooth noncommutative space. 
Its differential structure is most conveniently
 defined in the noncommutative frame formalism through momenta $\hp_\alpha$:
\begin{equation}
 d \hat f =(e_\alpha\hat f)\, \theta^\alpha =[\hp_\alpha, \hat f] \, \theta^\alpha \,.
\end{equation}
It is assumed that  frame  basis forms $\theta^\alpha$ 
commute with  functions on the algebra,
$[\hat f,\theta^\alpha] = 0 $. We choose the momenta as
\begin{equation}
 \epsilon \hp_1 =i \mu^2 \hx^2, \qquad \epsilon \hp_2 =-i\mu^2 \hx^1,\qquad 
\epsilon \hp_3 =i\mu (\mu \hx^3- \frac 12) .
\end{equation} 
With this definition the differential $d$ and the derivations $\, e_\alpha$, $\alpha = 1,2\,$
reduce to the differential and  the derivations on the Moyal plane
after the  dimensional reduction.

The U(1) gauge theory on the truncated Heisenberg space was constructed in \cite{Buric:2010xs}.
It is given by an  antihermitian  gauge potential $\hat A=\hat A_\alpha\theta^\alpha$; the field strength is
 $\hat \tF = d\hat A +\hat  A^2$. The commutation relations between the momenta define almost
completely the geometry; in this case they are quadratic
\begin{equation} 
 [\hp_\alpha,\hp_\beta] = \frac{1}{i\epsilon} K_{\alpha\beta} +F^\gamma{}_{\alpha\beta} \hp_\gamma
-2 i\epsilon Q^{\gamma\delta}{}_{\alpha\beta}\hp_\gamma \hp_\delta     ,       \label{quad}
\end{equation} 
with
\begin{equation}
 K_{12} =\frac {\mu^2}{2},\qquad F^1{}_{23} = \mu ,\qquad
 Q^{13}{}_{23} = \frac 12, \qquad  Q^{23}{}_{31} = \frac 12 .  \label{struc}
\end{equation}
This implies that the frame components of the field strength are 
\begin{equation}
 \hat \tF_{\alpha\beta} =\nabla_{[\alpha}\hat A_{\beta]} +[\hat A_\alpha,\hat A_\beta] +
 2i\epsilon (e_\eta\hat A_\gamma)Q^{\eta\gamma}{}_{\alpha\beta}+ 2i \epsilon\hat A_\eta\hat A_\gamma 
Q^{\eta\gamma}{}_{\alpha\beta} .
\end{equation}
The gravity-covariant derivative $\, \nabla_\alpha\hat A_\beta 
 = e_\alpha\hat A_\beta -\hat A_\gamma \hat\omega^\gamma{}_{\alpha\beta} \,      $
is here given through  connection $\hat \omega^\alpha{}_\beta$,
\cite{Buric:2009ss}:
\begin{eqnarray} 
 &&\hat \omega_{12} = -\hat\omega_{21} =  \mu\, 
(\frac 12-2\mu \hx^3)\, \theta^3 ,               \label{Con}\\
&&\hat \omega_{13} = -\hat\omega_{31}  = 
\frac \mu 2\, \theta^2 +2 \mu^2 \hx^1\,\theta^3 ,\nonumber \\
&&\hat \omega_{23} = -\hat \omega_{32}
= -\frac \mu 2 \, \theta^1 +2\mu^2 \hx^2\,\theta^3 .\nonumber
\end{eqnarray}
Instead in terms of  potential $\hat A_\alpha,$ the field strength can  be expressed in terms
of  covariant coordinates $\,\hat \tX_\alpha = \hp_\alpha +\hat A_\alpha\, $  as
\begin{equation}
\hat \tF_{\alpha\beta} = 2P^{\gamma\delta}{}_{\alpha\beta}{\hat\tX}_\gamma {\hat\tX}_\delta
 -F^\gamma{}_{\alpha\beta}{\hat\tX}_\gamma -\frac{1}{i\epsilon}K_{\alpha\beta} .                        \label{Falbe}
\end{equation} 
Covariant coordinate 1-form  $\,\hat\tX =\hat \tX_\alpha \theta^\alpha\, $ is 
a difference of two connections, 
$\hat A$ and the Dirac operator $\hat\theta=-\hp_\alpha\theta^\alpha$. It
transforms therefore in the adjoint representation of U(1), that is covariantly.

Performing the  dimensional reduction $\, \hx^3 =0$, the
 third component of the gauge field becomes a scalar, $\hat A_3 =\hat\phi$. The 
components of the field strength become
\begin{eqnarray}
&&\hat\tF_{12} =\hat F_{12}-\mu\hat \phi =[\hat\tX_1,\hat\tX_2] + \frac{i\mu^2}{\epsilon} -\mu\hat\phi ,      \label{field}        \\[6pt]
&&\hat\tF_{13} ={D}_1\hat\phi - i\epsilon \{\hp_2 + \hat A_2,\hat \phi\} = [\hat\tX_1,\hat\phi]
 -i\epsilon \{\hat \tX_2,\hat \phi\} ,   \nonumber  \\[10pt]
&&\hat\tF_{23} ={D}_2\hat\phi + i\epsilon \{\hp_1 +\hat A_1, \hat\phi\} = [\hat\tX_2,\hat\phi] +i\epsilon 
\{\hat \tX_1, \hat\phi \}\nonumber ,\\[-12pt]  \nonumber
\end{eqnarray}
where the gauge-covariant derivative is $D_\alpha\hat\phi =[\hp_\alpha +\hat A_\alpha,\hat \phi]\, $, $\alpha =1,2$,
and $\,\hat F =\hat F_{12}\,\theta^1\theta^2\, $ is a two-dimensional field strength,
$\,\hat F_{12} = \p_1\hat A_2 - \p_2\hat A_1 +[\hat A_1,\hat A_2]$.

Introducing (\ref{field}) into the Yang-Mills action we obtain
\begin{eqnarray}
\label{YM-curved}
 &&\cS_{YM}= \frac 12 \,\tr \Big( (1-\epsilon^2)(\hat F_{12})^2 -2 (1-\epsilon^2)\mu\hat F_{12}\hat\phi
 + (5 -\epsilon^2) \mu^2\hat \phi^2 + 4i\epsilon\hat F_{12}\hat \phi^2  \label{L}
\\
&&\phantom{S = \frac 12 \tr \quad\ } + (D_1\hat \phi)^2 +(D_2\hat \phi)^2 
-\epsilon^2\{ \hp_1 +\hat A_1,\hat \phi\}^2 -\epsilon^2\{ \hp_2 +\hat A_2,\hat \phi\}^2 \Big)  \nonumber
\end{eqnarray}
which defines our gauge model.
It is clear that  masses and couplings in (\ref{L}) are
fixed by the  dimensional reduction procedure
and parametrized  by only one parameter,  noncommutativity $\epsilon$. 
 Modifications of the  action (\ref{L}) are possible
but only at an earlier stage: for example, one can  use different 
connection $\hat \omega^\alpha{}_\beta$, or define the Hodge-dual differently.
Action (\ref{L}) has two stationary points: $\,\hat A_\alpha =0$, $\hat\phi =0\, $ and 
$\,\hat \tX_\alpha =0$, $\hat \tX_3=0$. Additional minima might
exist, but  the corresponding equations are quite complicated
and we were not able to find them in the generic case.

 It is  possible,  in the context of the truncated Heisenberg algebra,
 to  define the Chern-Simons action, too  \cite{Buric:2010xs}. It is given by
\begin{equation}
 \cS_{CS} = c\mu\, \tr\, 
\Big( (3-\epsilon^2)(\hat F_{12}-\frac{i\mu^2}{\epsilon})\,\hat \phi 
+\frac{2\ep}{3}\big( (\hp_1+\hat A_1)^2 + (\hp_2 +\hat A_2)^2\big) (\hat \phi -\frac{i\mu}{2\epsilon})
\Big) ,                                                         \label{modelCS}
\end{equation} 
where $c$ is an arbitrary constant. Adding $\, \cS_{CS}\,$ to  $\, \cS_{YM}\,$ 
changse of course the classical equations of motion: in principle,
 only the second vacuum, $\hat\tX_\alpha =0$, $\hat \tX_3=0\, $ remains. 
There are however special cases,  \cite{Buric:2010xs}.

\initiate
\section{Quantization}
\label{sec:quantization}

It was shown in \cite{Buric:2010xs} that one can introduce,
in a completely straightforward manner,  the gauge fixing and 
the ghost terms to (\ref{YM-curved}) and define a
 quantum action which is BRST invariant. We shall here 
 quantize the model perturbatively. To this end, we
 use the Moyal-space representation.
 This means that fields $\phi$, $A_\alpha$  are represented
by functions of commuting coordinates $x^\mu$, $\mu =1,2$,  while the
algebra-multiplication is represented by the Moyal-Weyl $\star$-product:
\begin{equation}
 \chi(x) \star \phi(x) = e^{\, \tfrac{i}{2}\, \theta^{\mu\nu}\p_\mu\p_\nu^\prime }
\chi(x) \phi(x^\prime) \vert_{x^\prime\to x}        ,                            \label{Moy}
\end{equation}
where in our case $ \theta^{\mu\nu} $ is
\begin{equation}
 \theta^{\mu\nu} = \frac{\epsilon}{\mu^2}\, \epsilon^{\mu\nu}.
\end{equation}
The signature is Euclidean. 
In the following we will often use  abbreviation
\begin{equation}
  \tilde x_\mu = \epsilon_{\mu\nu} x^\nu .
\end{equation}
Also, we will redefine antihermitian fields of the previous section to
 hermitian ones by
$A_\alpha\to iA_\alpha$, $\phi \to i\phi$. The classical
Yang-Mills action  is then written as
\begin{eqnarray}
&& \cS_{YM} =-  \frac 12 \,\int  a (F_{12})^{ 2} 
-2 a \mu F_{12}\star \phi + (4+a) \mu^2 \phi\star \phi 
+ 4i\epsilon F_{12}\star \phi\star \phi           ,        \label{modelM}
\\[2pt]
&&\phantom{ \cS_{YM} =-\int }
 + [  p_1 +A_1\ds \phi]^{ 2} +[  p_2 +A_2 \ds \phi ]^{ 2}
 -\epsilon^2\{ p_1 +A_1\ds \phi\}^{2} -\epsilon^2\{ p_2 +A_2 \ds \phi\}^{2}    \nonumber
\end{eqnarray}
where 
\begin{equation}
 a=1-\epsilon^2\,  .
\end{equation}
It can further  be simplified using the property
\begin{equation}
 \{ x^\mu\ds \phi\} = 2x^\mu \phi 
\end{equation}
and
\begin{equation}
 \p_\alpha\phi = [p_\alpha\ds \phi] , \qquad D_\alpha\phi = [p_\alpha+ A_\alpha\ds \phi] ,
\end{equation}
which is the form which we  use.
The gauge is fixed by a  non-covariant Lorentz gauge term:
\begin{equation}
 \cS_{gf} = \frac a2 \int (\p_\mu A^\mu )^2 .
\end{equation}
Other possibilities of the gauge fixing, discussed in \cite{Buric:2010xs}, result
much more complicated propagators.
Finally, we  add  the ghost term
\begin{equation}
 \cS_{gh} = -\int \bar c\p_\alpha \big(\p^\alpha c +i[A^\alpha \ds c]\big ).
\end{equation}
The quantum action is the sum of all three  terms,
\begin{equation}
\cS = \cS_{YM} + \cS_{gf} +\cS_{gh} = \cS_{kin} + \cS_{int}.
\end{equation}
It is BRST invariant; clearly we can add the Chern-Simons action
and retain this invariance.
The kinetic part of the action, after the gauge fixing, is
\begin{equation}
 \cS_{kin} =-\frac 12 \int~  a A_\alpha \Box A^\alpha + 2a\mu \epsilon^{\alpha\beta}(\p_\alpha A_\beta)\phi
+ \phi  \Box \phi - (4+a) \mu^2  \phi^2- 4\mu^4 x^\alpha x_\alpha \phi^2  + 2\bar c\, \Box c        \label{ki}
\end{equation}
whereas the  interaction reads
\begin{eqnarray} 
  &&\hskip-2cm\cS_{ int}=-\frac 12 \int~ 
4\epsilon \epsilon_{\alpha\beta}(\p^\alpha A^\beta +
i A^\alpha\star A^\beta)\star \phi^2 -2i(\p_\alpha\phi)[A^\alpha\ds \phi] \\[4pt]
&&\hskip-1cm\phantom{S_{ int}=}
+ 2ia\mu \epsilon_{\alpha\beta} A^\alpha\star A^\beta \phi 
-2ia \epsilon_{\alpha\beta}\p^\alpha A^\beta  \epsilon_{\gamma\delta}A^\gamma\star A^\delta 
+a(\epsilon_{\alpha\beta} A^\alpha\star A^\beta)^2\nonumber\\[8pt]
&&\hskip-1cm\phantom{S_{ int}=}
+[A_\alpha\ds \phi] [A^\alpha\ds \phi] -\epsilon^2 \{ A_\alpha\ds \phi\} \{ A^\alpha\ds \phi\} +2\mu^2\epsilon\epsilon_{\alpha\beta}\{ x^\alpha\ds\phi\} \{A^\beta\ds\phi\}
 \nonumber\\[8pt]
&&\hskip-1cm\phantom{S_{ int}=}
-i \bar c\p_\alpha[A^\alpha \ds c]\,.     \nonumber
\end{eqnarray}

\subsection{Propagators}

The first property of (\ref{ki}) which one observes is 
 that, although  the gauge fixing removes the mixed 
quadratic terms $\,(\p_\alpha A_\beta)( \p^\beta A^\alpha)$, 
the mixing between  $\phi$ and $ A_\alpha$  remains. This
 means that $A_\alpha$ and $\phi$ do not  propagate independently. 
Apparently, only the scalar
 $\phi$ is  coupled  to the oscillator potential $x^\alpha x_\alpha$. 
The value $a = 0$ of the parameter $a=1-\epsilon^2$
is special: for $a=0$ the gauge field does not propagate, that is the corresponding
kinetic term vanishes and  the model becomes  degenerate.  On the other hand
  $a=0$  would seem to be a preferred choice because 
for this value  representations of the truncated Heisenberg algebra
are finite matrices. We shall 
see that though some of the expressions which we will calculate much simplify
 for $a=0$, the structure of divergences is in fact similar for all values of $a$.

Since scalar and gauge field are mixed we consider them as a multiplet 
of fields, $(A^\mu ~ \phi)$.  The kinetic term can be rewritten as
\begin{equation}
 \cS_{kin} =-\frac 12 \int\, 
 \begin{pmatrix}  A^\mu  & \phi           \end{pmatrix}
\begin{pmatrix}
 a \Box \delta_{\mu\nu} &  -a \mu  \epsilon_{\mu\zeta} \p^\zeta    \\[8pt]
a \mu \epsilon_{\nu\eta} \p^\eta      &         K^{-1}-a\mu^2
\end{pmatrix}
\begin{pmatrix}    A^\nu \\[6pt]   \phi  \end{pmatrix}
 + 2\bar c\, \Box c ,
\end{equation}
where $K^{-1}$ denotes the operator
\begin{equation}
 K^{-1} =\Box - 4 \mu^4x_\alpha x^\alpha- 4 \mu^2 .
\end{equation}
The corresponding kinetic matrix
\begin{equation}
 G^{-1} = \begin{pmatrix}
a \Box \delta_{\mu\nu} &  -a \mu  \epsilon_{\mu\zeta} \p^\zeta    \\[8pt]
a \mu \epsilon_{\nu\eta} \p^\eta      &         K^{-1}-a\mu^2
 \end{pmatrix}   \\
\end{equation}
cannot be diagonalized easily because of the mixing of  position 
and  momentum variables introduced through  $K^{-1}$. It is however  possible to find 
its inverse, the propagator $G$:
\begin{equation}
G = \begin{pmatrix}
  \frac 1a \Box^{-1} \delta_{\mu\nu} -\mu^2 \Box^{-1}\epsilon_{\mu\zeta}\p^\zeta K
\epsilon_{\nu\eta} \p^\eta \Box^{-1} \ \  &   
-\mu \Box^{-1}\epsilon_{\mu\zeta}\p^\zeta  K     \\[8pt]
\mu K \epsilon_{\nu\eta}\p^\eta \Box^{-1}  &            K
 \end{pmatrix}   .\\                                             \label{propagator}
\end{equation}
The propagation of ghosts  is decoupled and the
 ghost propagator $G_{gh}$ is simple, $   G_{gh} ^{-1} =\Box$ .

We will use mostly the propagator  kernels in  momentum representation. 
In accordance with the conventions for the Fourier transformation
given in  Appendix I we have
\begin{eqnarray}
 && \tilde G_{gh} ^{-1}(p,q) = -(2\pi)^2 p^2 \delta(p+q)\,, \\[12pt]
&&\tilde G^{-1}(p,q)  = \begin{pmatrix}
- a p^2  \delta_{\mu\nu}(2\pi)^2 \delta(p+q) & - i a\mu  \epsilon_{\mu\zeta} p^\zeta  (2\pi)^2 \delta(p+q)   \\[8pt]
 -ia \mu  \epsilon_{\nu\eta} q^\eta  (2\pi)^2 \delta(p+q)   &  \tilde K^{-1}(p,q) -a\mu^2  (2\pi)^2 \delta(p+q)
 \end{pmatrix}  . \\ \nonumber
\end{eqnarray}
Note that in our conventions all momenta are incoming.
This will, apart from a somewhat unusual factor $\delta(p+q)  $
in the propagators, also reflect later when we define `long' and `short' variables.

 $K^{-1}$ is the kinetic operator for the  scalar field.
Its kernel in  coordinate space is given by
\begin{equation}
  K^{-1}(x,y) =  ( \Box -4\mu^4x_\alpha x^\alpha -4\mu^2 )\delta^2 (x-y) ;
\end{equation}
 the inverse is the so-called Mehler kernel. In two dimensions the Mehler kernel is
given by
\begin{equation}
 K(x,y) =-\frac{1}{8\pi} \int_0^\infty\frac{\omega\, d\tau}{\sinh \omega\tau}
e^{-\frac{\mu^2}{2}\big((x-y)^2\coth\frac{\omega\tau}{2} +(x+y)^2
\tanh\frac{\omega\tau}{2}\big)-\omega\tau},               \label{Kposit}
\end{equation}
or in  momentum space,
\begin{equation}
 \tilde K(p,q) =-\frac{\pi}{4 \mu^4} \int_0^\infty\frac{\omega\, d\tau}{\sinh \omega\tau}
e^{-\frac{1}{8\mu^2}\big((p+q)^2\coth\frac{\omega\tau}{2} 
+(p-q)^2\tanh\frac{\omega\tau}{2}\big)-\omega\tau}.                     \label{Komega}
\end{equation}
Parameter $\omega $ in the last formula is by dimension a frequency; it is in fact the
frequency of the quantum-mechanical harmonic oscillator of mass $m$
of which (\ref{Komega}) is the Green function. Here,  $m\omega = 2\mu^2$.
The mass  of the scalar field  on the other hand is $\mu_0$; it
 enters the Mehler kernel
 through  exponential term  $ \, e^{-\frac{\mu_0^2}{2m}\tau } =
e^{-\frac{\mu_0^2}{4\mu^2}\omega\tau}$. In
 our case this mass is fixed $\, \mu_0 = 2\mu \, $ and therefore we have 
in (\ref{Komega})  factor $ \, e^{-\omega\tau } $.

One usually introduces dimensionless parameter $\,\alpha = \omega\tau\,$ so the
Mehler kernel becomes
\begin{equation}
 \tilde K(p,q) =-\frac{\pi}{4 \mu^4} \int_0^\infty\frac{d\alpha}{\sinh \alpha}
e^{-\frac{1}{8\mu^2}\big((p+q)^2\coth\frac{\alpha}{2} +(p-q)^2\tanh\frac{\alpha}{2}\big)-\alpha},
\end{equation}
or  $\, \xi =\coth \frac{\alpha}{2}$ so we have
\begin{equation}
 \tilde K(p,q) =-\frac{\pi}{4 \mu^4} \int_1^\infty\frac{d\xi}{\xi}\, \frac{\xi -1}{\xi +1}
e^{-\frac{1}{8\mu^2}\big((p+q)^2\xi +(p-q)^2\frac{1}{\xi}\big)}.
\end{equation}
The importance of the form  (\ref{Komega}) of the Mehler kernel is that using it
one can easily perform  the free-field limit $\omega\to 0$. This limit, in particular, 
 determines the prefactors in kernels (\ref{Kposit}) and (\ref{Komega}): using the 
Schwinger parametrization and 
\begin{equation}
 \lim_{\sigma\to 0}\, \frac{1}{2\pi\sigma^2} \, e^{-\frac{(p+q)^2}{2\sigma^2}}=\delta^2(p+q) 
\end{equation}
we find the limiting value,
\begin{equation}
 \tilde K(p,q) \vert_{\omega\to 0} = -\frac{(2\pi)^2}{p^2+\mu_0^2}\,  \delta^2(p+q) .
\end{equation}

The Mehler kernel $K(x,y)$ is the contraction of two scalar fields in  coordinate space:
\begin{equation}
K(x,y) = \bcontraction{}{\phi(x)}{}{\phi(y)} \phi(x)\phi(y) 
= \frac{1}{(2\pi)^4} \int dk \,dl\, 
\bcontraction{}{\tilde \phi(k)}{}{\tilde \phi(l)} \tilde\phi(k)\tilde \phi(l) 
e^{-ikx-ily}\, ;
\end{equation}
 in  momentum space we write
\begin{equation}
\bcontraction{}{\tilde \phi(k)}{}{\tilde\phi(l)} \tilde \phi(k)\tilde \phi(l) = \tK(k,l) .
\end{equation}
In  the following,  in order  to alleviate  the notation
we will omit the tilde sign in the Fourier transformation,
so we will distinguish for example $\phi(x)$ from $\tilde\phi(p)$ by 
the value of the argument only. We thus write
\begin{equation}
\bcontraction{}{\tilde \phi(k)}{}{\tilde \phi(l)} \tilde \phi(k) \tilde \phi(l)  
\equiv \bcontraction{}{\phi(k)}{}{\phi(l)} \phi(k) \phi(l) = K(k,l) ,     \label{28}
\end{equation}
and in analogy
\begin{eqnarray}
&&  \bcontraction{}{A_\sigma(k)}{}{\phi(l)} A_\sigma (k) \phi(l)  = - i \mu \,  
\frac{\epsilon_{\sigma\beta} k^\beta}{k^2}\, K(k,l) ,\\[4pt]
&& \bcontraction{}{\phi(k)}{}{A_\sigma(l)} \phi(k) A_\sigma(l) = -i \mu K(k,l) \, \frac{\epsilon_{\sigma\beta} l^\beta}{l^2}
,\\[4pt]
&&  \bcontraction{}{A_\rho(k)}{}{A_\sigma(l)} A_\rho(k) A_\sigma(l) =-  \frac {(2\pi)^2 }a \frac{\delta_{\rho\sigma}}{k^2}\,  \delta(k+l)
+(-i\mu)^2 \, \frac{\epsilon_{\rho\nu}k^\nu}{k^2}\, K(k,l) \, \frac{\epsilon_{\sigma\tau} l^\tau}{l^2}\,,
\\[4pt]
&&  \bcontraction{}{\bar c(k)}{}{c(l)} {\bar c} (k) c(l) = -  \frac {(2\pi)^2 }{k^2}\,  \delta(k+l)\,. \label{32}
\end{eqnarray}

\subsection{Vertices}

Transforming the interaction terms to 
 momentum space we obtain the following 3-vertices:
\begin{eqnarray*}
&&  (1)\quad
%         -2\epsilon \int (\p_1 A_2 -\p_2 A_1)\star\phi\star\phi  \ \to\ \\[6pt]
% &&\phantom{\qquad\qquad}
\frac {2i\epsilon}{(2\pi)^4}\int dp\,dq\, dk\,\delta(p+q+k)\cos \frac{k\wedge q}{2}\, \epsilon_{\rho\sigma}
p^\rho A^\sigma(p)\phi(q)\phi(k) 
\hskip4cm \\[8pt]&&  (2)\quad
% i\int (\p_1\phi)[A_1\ds \phi] + (\p_2\phi) [A_2\ds \phi] \ \to\ \\[6pt]
% &&\phantom{\qquad\qquad}
\frac {2i}{(2\pi)^4}\int dp\,dq\, dk\,\delta(p+q+k) \sin \frac{q\wedge k}{2}\, p^\rho \phi(p) 
 A_\rho(k)\phi(q)
\\[8pt]&&  (3)\quad
% \mu^2 \epsilon \int \{x^2\ds \phi\} \{A_1\ds \phi\} -\{x^1\ds \phi\} \{ A_2\ds \phi\} \ \to \ \\[6pt]
% &&\phantom{\qquad\qquad}
\frac {-4i\mu^2\epsilon}{(2\pi)^4}\int dp\,dq\, dk\,\delta(p+q+k) \cos \frac{k\wedge q}{2} 
\epsilon_{\rho\sigma}\frac{\p\phi(p)}{\p p_\sigma}A^\rho(k) \phi(q)
\\[8pt]&&  (4)\quad
% -ia \mu \int[A_1\ds A_2] \phi \ \to\ \\[6pt]
% &&\phantom{\qquad\qquad}
\frac {a\mu}{(2\pi)^4}\int dp\,dq\, dk\,\delta(p+q+k) \sin \frac{q\wedge p}{2}
\epsilon_{\rho\sigma}A^\rho(p) A^\sigma(q) \phi(k)
\\[8pt]&&  (5)\quad
% ia \int (\p_1A_2-\p_2 A_1)[A_1 \ds A_2] \ \to\ \\[6pt]
% &&\phantom{\qquad\qquad}
\frac {ia}{(2\pi)^4}\int dp\,dq\, dk\,\delta(p+q+k) \sin \frac{q\wedge p}{2}
\epsilon_{\rho\sigma} \epsilon_{\lambda\tau}k^\lambda A^\rho(p) A^\sigma(q)  A^\tau(k) \\
[8pt]&&  (6)\quad
% -i \int \bar c  \p_\alpha [A^\alpha \ds c]   \ \to\ \\[6pt]
% &&\phantom{\qquad\qquad}
\frac {2i}{(2\pi)^4}\int dp\,dq\, dk\,\delta(p+q+k) \sin \frac{q\wedge k}{2}\,
p_\alpha \bar c(p) c(q) A^\alpha(k)\,,    \\   \nonumber
\end{eqnarray*}
where we  denoted
\begin{equation}
 p \wedge q = \frac{\epsilon}{ \mu^{2} }\,  \epsilon^{\alpha\beta}p_\alpha q_\beta 
= \frac{\epsilon}{ \mu^{2} }\,  p\cdot \tilde q  .
\end{equation}
Clearly only the vertex (3),  containing the derivative of  $\delta$-function, comes
from the position-dependent terms in the interaction and  breaks the translation invariance.
There are no such terms in  4-vertices as we have:
\begin{eqnarray*}
&&  (7)\quad
% -\frac 12 \int[A_1\ds \phi]^2 +[A_2\ds\phi]^2 \ \to
% \\[6pt]
% &&\phantom{\quad\quad}
\frac{2}{(2\pi)^6}\int dp\, dq\, dk\, dl \, \delta(p+q+k+l) \sin\frac{k\wedge p}{2}
 \sin\frac{l\wedge q}{2}\delta^{\rho\sigma} A_\rho(p) A_\sigma(q)\phi(k)\phi(l)
\\[8pt]&&  (8)\quad
% \frac {\epsilon^2}{2} \int \{ A_1\ds \phi \} ^2 +\{ A_2\ds\phi\} ^2 \ \to
% \\[6pt]
% &&\phantom{\quad\quad}
\frac{2\epsilon^2}{(2\pi)^6}\int dp\, dq\, dk\, dl \, \delta(p+q+k+l) \cos\frac{k\wedge p}{2}
 \cos\frac{l\wedge q}{2}\delta^{\rho\sigma} A_\rho(p) A_\sigma(q)\phi(k)\phi(l)
\\[8pt]&&  (9)\quad
% -2i\epsilon \int [A_1\ds A_2]\star \phi^2 \ \to
% \\[6pt]
% &&\phantom{\quad\quad}
\frac{2\epsilon}{(2\pi)^6}\int dp\, dq\, dk\, dl \, \delta(p+q+k+l) \sin\frac{q\wedge p}{2}
 \cos \frac{l\wedge k}{2}\, \epsilon^{\rho\sigma} A_\rho(p) A_\sigma(q)
\phi(k)\phi(l)  \\[8pt]
&&  (10)\quad
% -\frac a2 \int [A_1\ds A_2]^2 \ \to
% \\[6pt]
% &&\phantom{\quad\quad}
\frac{a}{2(2\pi)^6}\int dp\, dq\, dk\, dl \, \delta(p+q+k+l) \sin\frac{q\wedge p}{2}
 \sin\frac{l\wedge k}{2}\epsilon^{\rho\sigma} A_\rho(p) A_\sigma(q)
\epsilon^{\lambda\tau} A_\lambda(k) A_\tau(l)  .
\end{eqnarray*}
This completes the list of the Feynman rules of our theory.

\initiate

\section{Tadpoles: one-loop divergences}

We start the calculation of the quantum corrections from the simplest,  tadpole diagram.
Since we have  vertices with three external lines such diagrams
 a priori exist. Moreover, as the translation invariance is broken and the
momentum is not conserved along the propagator, they do not vanish.
The scalar field tadpole is the expectation value
\begin{equation}
 T_{\phi}\equiv T(r) = -\langle \phi(r) \, \cS_{int}\rangle  .
\end{equation}
Nonvanishing contributions can be graphically represented
 as in the picture. It is worth stressing that  propagators
are drawn either by a simple flat line or by a mixed line,  corresponding to
respectively  $\bcontraction{}{\phi}{}{\phi}\phi\, \phi$ and $\bcontraction{}{\phi}{}{A_\mu}\phi \, A_\mu$.
\begin{center}
\includegraphics[scale=2]{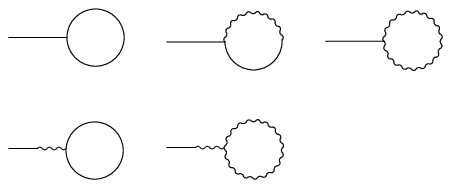}
\end{center}
We need contributions  from each of the 3-vertices.
  From  vertex $(1)$  we obtain
\begin{eqnarray}
 &&T_{\phi,1} =
\frac {2i\epsilon}{(2\pi)^4}\int dp\,dq\, dk\,\delta(p+q+k)\cos \frac{k\wedge q}{2}\, \epsilon_{\rho\sigma}
p^\rho \\[8pt]
&&\phantom{T_{\phi,1} = {2i\epsilon}}
\times\Big( \bcontraction{}{\phi(r)}{}{A^\sigma(p)} \phi(r) A^\sigma(p) \,  
\bcontraction{}{\phi(q)}{}{\phi(k)} \phi(q)\phi(k) +
\bcontraction{}{\phi(r)}{}{\phi(q)} \phi(r)\phi(q) \, \bcontraction{}{A^\sigma(p)}{}{\phi(k)} A^\sigma(p)\phi(k) +
\bcontraction{}{\phi(r)}{}{\phi(k)} \phi(r) \phi(k) \, 
\bcontraction{}{A^\sigma(p)}{}{\phi(q)} A^\sigma(p)\phi(q) \Big)   
\nonumber \\[8pt]
&&\phantom{T_{\phi,1} }=        \nonumber
\frac {2\mu\epsilon}{(2\pi)^4}\int dp\,dq\, dk\,\delta(p+q+k)\cos \frac{k\wedge q}{2}\, \cK(r,p,q,k) ,
\end{eqnarray}
where we  introduced the cyclic product of two Mehler kernels
\begin{equation}
 \cK(r,p,q,k) = K(r,p)K(q,k) +  K(r,q)  K(p,k) + K(r,k) K(p,q) .
\end{equation}
Obviously,  $\cK\,$ is invariant under  permutations of its factors which is similar to the property
\begin{equation}
 K(p,q) = K(q,p) = K(-p,-q)
\end{equation}
of the Mehler kernel. Analogously the other 3-vertices give
\begin{eqnarray}
 && T_{\phi,2} =-\frac {2\mu}{(2\pi)^4}\int dp\,dq\, dk\,\delta(p+q+k)\sin \frac{k\wedge q}{2}\, 
\frac{p\cdot\tilde k}{k^2}\, \cK(r,p,q,k) \\[6pt]
&& T_{\phi,3} =\frac {4\mu^3\epsilon}{(2\pi)^4}\int dp\,dq\, dk\,\delta(p+q+k)\cos \frac{k\wedge q}{2}\, 
\frac{ k_\sigma}{k^2}\, \frac{\p}{\p p_\sigma}\, \cK(r,p,q,k)  \nonumber \\[6pt]
&& T_{\phi,4} =\frac {a\mu^3}{(2\pi)^4}\int dp\,dq\, dk\,\delta(p+q+k)\sin \frac{q\wedge p}{2}\, 
\frac{p\cdot\tilde q}{p^2 q^2}\, \cK(r,p,q,k) \nonumber \\[6pt]
&& T_{\phi,5} = -\frac {a\mu^3}{(2\pi)^4}\int dp\,dq\, dk\,\delta(p+q+k)\sin \frac{q\wedge p}{2}\, 
\frac{p\cdot\tilde q}{p^2 q^2}\, \cK(r,p,q,k)\,. \nonumber \\[10pt]
&& T_{\phi,6} = 0.  \nonumber
\end{eqnarray}
The ghost contribution is zero, $\, T_{\phi,4} + T_{\phi,5}=0$ and  4-vertices
do not contribute. 
Therefore for the scalar-field tadpole we obtain:
\begin{equation}
 T(r) =\frac{2\mu}{(2\pi)^4}\int dp\,dq\, dk\,\delta(p+q+k)
\big(     \epsilon \cos \frac{p\wedge q}{2}\, (1 + 2\mu^2 \frac{p_\sigma}{p^2}\,\frac{\p}{\p q_\sigma})   
+\sin \frac{q\wedge p}{2}\, 
\frac{p\cdot\tilde q}{p^2 q^2}  \big)\, \cK(r,p,q,k) .                \label{TP}
\end{equation}
In  similar manner we can calculate the gauge-field tadpole,
\begin{equation}
 T_{A_\mu}\equiv T_\mu(r) = -\langle A_\mu(r) \, \cS_{int}\rangle =\sum_{j=1}^6  T_{A_\mu,j} ;
\end{equation}
the result is
\begin{eqnarray}
 &&T_{\mu} (r)=-\frac{2i\mu^2}{(2\pi)^4}\frac{\tilde r_\mu}{r^2}\int dp\,dq\, dk\,\delta(p+q+k)          \label{TA}\\
&&\phantom{T_{A_\mu}=\quad}
\times \Big( \epsilon \cos \frac{p\wedge q}{2}\, (1 + 2\mu^2 \frac{p_\sigma}{p^2}\,\frac{\p}{\p q_\sigma})   
+\sin \frac{q\wedge p}{2}\, 
\frac{p\cdot\tilde q}{p^2 q^2}  \Big)\, \cK(r,p,q,k) \nonumber \\[4pt]
&&\phantom{T_{A_\mu}=}
+\frac{i}{(2\pi)^2a}\, \int dp\,dq\, \delta(p+q-r)\nonumber\\
&&\phantom{T_{A_\mu}=\quad}
\times\Big( 2\epsilon \cos \frac{p\wedge q}{2}\,\frac{\epsilon_{\mu\alpha}}{r^2}(r^\alpha-2\mu^2\frac{\p}{\p p_\alpha}) 
+\sin \frac{q\wedge p}{2}\, (\frac{2 p_\mu}{r^2}+ a\mu^2 \frac{\tilde r_\mu}{r^2}
\frac{p\cdot\tilde q}{p^2 q^2} ) \Big)\, K(p,q)  . \nonumber
\end{eqnarray}
One notices that there
 is a simple relation betweenthe two expressions, (\ref{TP}) and (\ref{TA}):
\begin{equation}
 T_{\nu}(r) = -i\mu\, \frac{\tilde r_\nu}{r^2}\,  T (r) + \bbB_\nu(r) 
\end{equation}
where
\begin{eqnarray}
&&\bbB_\mu(r) =
 \frac{i}{(2\pi)^2a}\, \int dp\,dq\, \delta(p+q-r)
\\[4pt] &&\phantom{\bbB_\mu(r) =} \times
\Big( 2\epsilon \cos \frac{p\wedge q}{2}\,\frac{\epsilon_{\mu\alpha}}{r^2}(r^\alpha-2\mu^2\frac{\p}{\p p_\alpha}) 
+\sin \frac{q\wedge p}{2}\, (\frac{2 p_\mu}{r^2}+ a\mu^2 \frac{\tilde r_\mu}{r^2}\,
\frac{p\cdot\tilde q}{p^2 q^2} ) \Big)\, K(p,q)       . \nonumber
\end{eqnarray}

In fact, as we are using the multiplet of fields $\, (A_\mu\, \phi)\, ,$ it is more practical to
 change diagrammatic representation ofthe  propagators and use one, `doubled' line
for both fields in the multiplet as given below. Then the tadpole is also a doublet:
the corresponding diagram is
\begin{center}
\hskip-5cm
\includegraphics[scale=2]{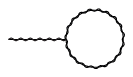}
\hskip1cm
 $   \begin{pmatrix}
  T_\mu(r)\\[4pt]  T(r)
 \end{pmatrix} $ .
\end{center}

Though  we have the results for the tadpole,
it is not easy to understand the structure of divergences in the obtained expressions.
This is difficult first of all because we are dealing with the Mehler kernel
in which results are in the form of a  parameter integral. In addition, 
 (\ref{TP}-\ref{TA})  contain $\,\cK(r,p,q,k)$,  that is the products 
of two kernels. In order to find
divergences and the form of counterterms we apply the method
developed in \cite{Blaschke:2009aw,Erw}: we  amputate the leg of the tadpole,
 multiply it with the corresponding external field  then and integrate.
The amputated tadpole graph is obtained by multiplication with the
 inverse propagator:
\begin{center}
 \includegraphics[scale=2]{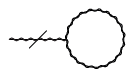} \hskip1cm
$ \begin{pmatrix}
  \tau_\mu(s)\\[4pt]   \tau(s) 
 \end{pmatrix}
= \dfrac{1}{(2\pi)^2}\displaystyle{ \int} dr \, G^{-1}(s, -r)
\begin{pmatrix}
 T_\nu(r)\\[4pt]  T(r)
\end{pmatrix} $ ;
\end{center}
 from (\ref{TP}-\ref{TA})  we get
\begin{eqnarray}
 && \tau_\mu(s) = -as^2 \bbB_\mu(s)\,, \\[4pt]
&& \tau(s) =ia\mu \tilde s^\nu  \, \bbB_\nu(s) + \frac{1}{(2\pi)^2} \int dr\,K^{-1}(s,-r)\, T(r) .
\end{eqnarray}

\subsection{Counterterms}

It is clear that our formulae simplify  considerably for $\,a=0$, so let us
 calculate first this part of the tadpole  divergences. 
This will help us to understand better the framework we are working in;  the
$a$-linear terms we will calculate in the sequel. After  momentum 
integrations we have
\begin{eqnarray}
 &&\tau_\mu(s)\vert\equiv \tau_\mu(s)\vert_{a=0} = \frac{i\epsilon}{4\mu^2}\, \tilde s_\mu  \int_1^\infty 
d\xi\, (\xi -1) \, e^{-\frac{s^2}{4\mu^2}\xi}          \label{tnu} \\[4pt]
&&\tau(s)\vert\equiv \tau(s)\vert_{a=0} = -\frac{\epsilon}{2\mu} \int_1^\infty d\xi\, \, \frac{\xi -1}{\xi +1}\, e^{-\frac{s^2}{4\mu^2}\xi}.
\end{eqnarray}
Both integrals  are finite in  variable $\xi$ but the result is 
divergent in external momentum $s$ in the infrared region, $s=0$:
\begin{eqnarray}
 &&\tau_\mu(s)\vert =4i\mu^2\frac{ \tilde s_\mu}{s^4} 
 \, e^{-\frac{s^2}{4\mu^2}}  \\[4pt]
&&\tau(s)\vert = -\frac{1}{\mu} \, e^{\frac{\, s^2}{4\mu^2}}\Big(E_0(\frac{s^2}{2\mu^2})
 - E_1(\frac{s^2}{2\mu^2})\Big) ,
\end{eqnarray}
where $E_0$ and $E_1$ are the exponential integrals reviewed shortly in  Appendix I.
As mentioned before, the corresponding counterterms can be found by multiplying  by  
external field and integrating. As
the  tadpoles are divergent  only at $s=0$  we can Taylor-expand
the external fielsd around this value and  integrate term by term.
For the gauge field we obtain\footnote{ In  this subsection  we 
reintroduce the tilde to distinghuish between a field and its Fourier transform,
that is to clarify the form of counterterms both in  momentum and in position space.}
\begin{eqnarray}
 &&\frac{1}{(2\pi)^2}\int d^2s\, {\tilde A}^\mu(s) \tau_\mu(-s)\vert =- \frac{4i\mu^2}{(2\pi)^2}\int
d^2s\, \frac{\tilde s_\mu}{s^4} e^{-\frac{s^2}{4\mu^2}} {\tilde A}^\mu(s)   \label{11}
 \\[8pt]
&&= -\frac{4i\mu^2}{(2\pi)^2}\int d^2s\,
\frac{\tilde s_\mu}{s^4} e^{-\frac{s^2}{4\mu^2}}\Big( {\tilde A}_\mu(0)+
\frac{\p {\tilde A}_\mu}{\p s_\rho}(0)\, s_\rho  +\frac{1}{2!}\, 
 \frac{\p^2 {\tilde A}_\mu }{\p s_\rho\p s_\sigma}(0)\, s_\rho s_\sigma
 +\dots \Big) .  \nonumber
\end{eqnarray}
Here and in the following, as we shall see, only the initial terms are divergent. In  (\ref{11})
the first integral is obviously zero while the second one can be calculated in polar coordinates,
$\, s^1 = s\cos\varphi$, $s^2 = s\sin\varphi$, $d^2 s = s \, ds d\varphi$. 
Using 
\begin{equation}
 \int_0^{2\pi} \frac{s_\alpha s_\beta}{s^2} \, d\varphi =\pi \delta_{\alpha\beta},
\end{equation}
we obtain
\begin{eqnarray}
 && \frac{1}{(2\pi)^2}\int d^2s\, {\tilde A}^\mu(s) \tau_\mu(-s)\vert =
 - \frac{4i\mu^2}{(2\pi)^2}\, \epsilon_{\mu\alpha}\frac{\p {\tilde A}_\mu}{\p s_\rho}(0) \,
\pi \delta_{\alpha\rho}\int_0^\infty  \frac{ds}{s}\, e^{-\frac{s^2}{4\mu^2}}
\nonumber     \\[8pt]
&&\phantom{\ \ \ \ }
=  - \frac{i\mu^2}{2\pi}\, \epsilon_{\mu\rho}\frac{\p {\tilde A}_\mu}{\p s_\rho}(0) \, \Gamma(0)
= -  \frac{\mu^2}{2\pi}\, \Gamma(0) \int d^2x\, \tilde x_\mu \tilde A^\mu (x) .
\end{eqnarray}
The integral has a  logarithmic divergence.  Higher than linear terms in 
the expansion contain higher orders of $s$ and therefore they converge
at the lower bound;  convergence at $s=\infty$ is 
 guaranteed by the exponentially  decreasing factor $ \,  e^{-\frac{s^2}{4\mu^2}} \ $.

In  similar way we can calculate the divergence  in $\tau(s)\vert$. As before,
the infinite contribution  comes  from the lower bound,  $s=0$: we will  focus
therefore on the behavior of the integral only at this point.
Expanding  \,$\tilde\phi(s)$ and the exponential integrals
$\,E_0(\frac{s^2}{2\mu^2})$ and $\, E_1(\frac{s^2}{2\mu^2})$ around zero,
we obtain
\begin{eqnarray}
&& \frac{1}{(2\pi)^2}\int d^2s\,\tilde \phi(s)\tau(-s)\vert =        \label{22}      \\[4pt]
&& \phantom{\quad} =
 - \frac{1}{\mu (2\pi)^2}\int d^2 s\,
e^{\frac{s^2}{4\mu^2}}\Big(\frac{2\mu^2}{s^2}\, e^{-\frac{s^2}{2\mu^2}} 
+\gamma + \log\frac{s^2}{2\mu^2} +\dots \Big) \Big( \tilde\phi(0) +  \frac{\p\tilde \phi}{\p s_\alpha}(0)\,
s_\alpha +\dots\Big)    . \nonumber
\end{eqnarray}
The first, divergent, term gives\footnote{As in the previous calculation we use 
$$ \hskip-4cm \int d^2x\,\phi(x)  =\frac{1}{(2\pi)^2}\iint d^2p\, d^2x\, \tilde\phi(p) e^{-ipx} =
\frac{1}{(2\pi)^2}\int d^2p\,\tilde\phi(p) (2\pi)^2\delta ^2(p)   =\tilde\phi(0) .
$$}
\begin{eqnarray}
 - \frac{\mu}{\pi}\, \int_0^\infty 
\frac{sds}{s^2}\, e^{-\frac{s^2}{2\mu^2}} \,\tilde\phi(0) 
=-\frac{\mu}{2\pi}\Gamma(0)\tilde\phi(0) =
 -\frac{\mu}{2\pi} \Gamma(0) \int d^2 x\, \phi(x)   ,            \label{Tt}
\end{eqnarray}
and the divergence is again logarithmic.
Other terms in (\ref{22})  give finite contributions including the $\log$ term, 
as the integral of the logarithm vanishes at the lower bound,  $
 \int \log s \, ds =s \log s -s \, $.

We obtained for $a=0$ only two counterterms which regularize the tadpole diagrams at one loop:
\begin{equation}
\int d^2x\, \phi,\qquad \int d^2x\, \tilde x^\mu \star A_\mu      .         \label{tadpole}
\end{equation}
 Are  there more counterterms in linear order in $a$? In fact
the answer is negative: though the integrals
which one calculates become more complicated, they do not
bring additional divergences. To see this we start again with $\, \tau_\mu$  denoting
\begin{equation}
 \tau_\mu(s) =\tau_\mu(s) \vert + \Delta\tau_\mu(s) ;
\end{equation}
 the difference is
\begin{equation}
\Delta\tau_\mu(s) = -a\,\frac{i\mu^2}{(2\pi)^2}\, \tilde s_\mu
\int dp\, dk\, \delta(p+k-s)\sin\,\frac{p\wedge k}{2}\, \frac{p\cdot\tilde k}{p^2 k^2}\, K(p,k). 
\end{equation}
The additional integral can be expressed in  parameter form using Schwinger 
parametrization. After momentum integrations  we obtain
\begin{eqnarray}
 &&\hskip-1cm
\Delta\tau_\mu =\frac{i\epsilon a}{4}\, \tilde s_\mu \int_1^\infty d\xi\, \frac{\xi (\xi -1)}{\xi +1}
 \int_0^\infty 
\frac{d\beta}{(1+4\mu^2\beta\xi)^2-\epsilon^2\xi^2}
e^{-\frac{s^2}{8\mu^2} \big(8\mu^2\beta +\xi +\frac {1}{\xi} 
-\frac{(1+4\mu^2\beta\xi)^2-\epsilon^2\xi^2}{\xi(1+2\mu^2\beta\xi) }\big)   }
 \nonumber        \\[8pt]
&&\hskip-1cm
\phantom{ \Delta\tau_\mu }=\frac{i\epsilon a}{8\mu^2}\, \tilde s_\mu
 \iint_1^\infty d\xi\, d\eta\, \frac{\xi -1}{\xi +1}\,
\frac{1}{(2\eta -1 -\epsilon \xi)(2\eta -1 +\epsilon \xi)} \,
e^{-\frac{s^2}{8\mu^2} \big(\xi +\frac {1}{\xi} 
+\frac{1}{\eta}(\epsilon^2\xi - \frac{1}{\xi})\big)   }   ,        \label{111}
\end{eqnarray}
where we introduced a new variable $\,\eta = 1 + 2\mu^2\xi\beta$.
Similarly for the  scalar-field tadpole 
$\,\Delta \tau (s)= \tau (s)-\tau (s)\vert_{a=0}\,$  we have
\begin{eqnarray}
 &&\hskip-1.5cm
 \Delta\tau(s) =\frac{a\epsilon}{8\mu^2}\,\iint_1^\infty  d\xi\, d\eta\, \frac{\xi -1}{\xi +1}\,
\frac{1}{(2\eta -1 -\epsilon \xi)(2\eta -1 +\epsilon \xi)} \,
e^{-\frac{s^2}{8\mu^2} \big(\xi +\frac {1}{\xi} 
+\frac{1}{\eta}(\epsilon^2\xi - \frac{1}{\xi})\big)   }   \nonumber          \\[8pt]
&&\hskip-1.5cm
 \phantom{ \Delta\tau(s) =}
+\frac{2a\epsilon}{\mu}\,e^{\frac{s^2}{8\mu^2} (1+\epsilon^2) }
 \int_1^\infty d\zeta \, (\frac{2}{\zeta} -\frac{1}{\zeta^2})
e^{-\frac{s^2}{4\mu^2} (1+\epsilon^2)\xi   } .             \label{222}
\end{eqnarray}
The second line of (\ref{222})  can be integrated in terms of the exponential integrals and 
it is convergent for all $s$. The double integral is the same for
 both corrections (\ref{111}) and (\ref{222}): it is regular, too. This we can verify by
analyzing the integral in the potentially divergent
 region $s=0$, in which the exponential can be replaced by 1. The integral is then
\begin{eqnarray}
 && \iint_1^\infty d\xi\, d\eta \, \frac{\xi -1}{\xi +1} \,
\frac{1}{(2\eta -1)^2 -\epsilon^2\xi^2} =\frac12\,   \iint_1^\infty  d\xi\, d\zeta \, 
\frac{\xi -1}{\xi +1} \,
\frac{1}{\zeta ^2 -\epsilon^2\xi^2}        \nonumber \\[8pt]
&&  =
\frac{1}{4\epsilon}\,   \int_1^\infty \frac{ d\xi}{\xi} \, \frac{\xi -1}{\xi +1} \,
\log\vert\frac{1-\epsilon\xi}{1 +\epsilon\xi}\vert ,
\end{eqnarray}
and  finite for all values of $\epsilon$. For $\epsilon=1\,$ 
for example the value of the integral is $\, -\frac{\pi^2}{24}\,$
while for general $\epsilon$ it obtains additional terms  proportional
to the PolyLog functions. We thus arrive at a very nice conclusion, that there are no 
new divergences in the tadpole  diagrams for $a\neq 0$ apart from those 
 given in (\ref{tadpole}).

\initiate
\section{Propagators: one-loop divergences}

The one-loop  propagator corrections can be calculated along similar lines
except that  calculations are longer and more complicated.  We denote
\begin{eqnarray}
&& P_{\phi(r)\phi(s)}\equiv P(r,s) =-\langle  \phi(r)\phi(s) \cS_{int}\rangle, \\[4pt]
&& P_{\phi(r)A_\mu(s)}\equiv P_\mu(r,\mu s) =  P_\mu(\mu s,r) 
=-\langle  \phi(r)A_\mu(s) \cS_{int}\rangle ,
\nonumber \\[4pt]
&& P_{A_\nu(r)A_\mu(s)} \equiv P_{\nu\mu}(\nu r,\mu s)=-\langle A_\nu(r)A_\mu(s) \cS_{int}\rangle .
\nonumber
\end{eqnarray}
The one-loop  corrections contain now three field contractions so we introduce auxilliary 
functions
\begin{eqnarray}
&& N(r,s,p,q,k,l) = K(r,s)K(k,l)K(p,q) + {\rm all \ pairings\ of \ arguments} \\[4pt]
&&\phantom{N(r,s,p,q,k,l)}
\equiv \cN (r,s;p,q,k,l) +K(r,s)\,\cK (p,q,k,l) . \nonumber
\end{eqnarray}
The product of three Mehler kernels
 $ N(r,s,p,q,k,l) $ contains 15 terms that is,  all permutations of the
arguments. The $\cN (r,s;p,q,k,l) $ on the other hand
 does not change under permutations of the first two 
and of the last four. To simplify  expressions for the 2-point
Green  functions we introduce further:
\begin{eqnarray}
 && \mathbb{N}(r,s) 
%=\int dp \,dq\,dk\,dl \, \, \delta (p+q+k+l) \, \cN (rs,pqkl)   \nonumber \\[8pt]
% &&\phantom{{\mu^2}{(2\pi)^6} }
% \times  \Big(  2 \, \frac{p\cdot q}{p^2 q^2} \,(\, \sin \frac{k\wedge p}{2}\, \sin \frac{l\wedge q}{2} 
% +\epsilon^2  \cos \frac{k\wedge p}{2}\, \cos\frac{l\wedge q}{2} )
%  + \frac 12 \,\frac{\epsilon^{\alpha\beta}p_\alpha q_\beta}{p^2q^2} \, \sin \frac{q\wedge p}{2}\,
% \cos \frac{l\wedge k}{2}  ) \nonumber \\[8pt]
% &&\phantom{{\mu^2}{(2\pi)^6} \times }
%  - \frac{a\mu^2}{2}\, 
% \frac{\epsilon^{\alpha\beta}p_\alpha q_\beta}{p^2q^2}\,\,
%  \frac{\epsilon^{\gamma\delta} k_\gamma l_\delta}{k^2 l^2} \,
% \sin \frac{q\wedge p}{2}\, \sin \frac{l\wedge k}{2} \,
% \Big)  
%  \nonumber \\[8pt]
% &&\phantom{{\mu^2}{(2\pi)^6} }
=\int dp \,dq\,dk\,dl \, \, \delta (p+q+k+l) \, \cN (r,s;p,q,k,l)   \nonumber \\[8pt]
&&\phantom{{\mu^2}{(2\pi)^6} }
\times  \Big(  2 \, \frac{p\cdot q}{p^2 q^2} \,(\, \sin \frac{q\wedge p}{2}\, \sin \frac{l\wedge k}{2} 
+  \cos \frac{q\wedge p}{2}\, \cos\frac{l\wedge k}{2} )
 + \frac 12 \,\frac{p\cdot\tilde q}{p^2q^2} \, \sin \frac{q\wedge p}{2}\,
\cos \frac{l\wedge k}{2}   \nonumber \\[8pt]
&&\phantom{{\mu^2}{(2\pi)^6} \times }
 - a\, \big( 2 \, \frac{p\cdot q}{p^2 q^2} \, \cos \frac{k\wedge p}{2}\, \cos\frac{l\wedge q}{2}
+ \frac{\mu^2}{2}\, 
\frac{p\cdot\tilde q}{p^2q^2}\,\,
 \frac{ k \cdot \tilde l}{k^2 l^2} \,
\sin \frac{q\wedge p}{2}\, \sin \frac{l\wedge k}{2} \big)
\Big)
\\[16pt]
 && \bbK_\nu (s,r)
%=\int dp \,dk\,dl \, \, \delta (p+k+l-r) \, \cK(spkl)  \nonumber \\[8pt]
% &&\phantom{{\mu^2}{(2\pi)^6} }
% \times\Big( 4 \,  \frac{\epsilon_{\nu\alpha}p^\alpha}{p^2} \,(-\sin \frac{k\wedge p}{2}\, \sin \frac{l\wedge r}{2} 
% +\epsilon^2  \cos \frac{k\wedge p}{2}\, \cos\frac{l\wedge r}{2} )
% - \frac{p_\nu}{p^2} \, \sin \frac{r\wedge p}{2}\,
% \cos \frac{l\wedge k}{2} 
% \nonumber\\[8pt]
% &&\phantom{{\mu^2}{(2\pi)^6}\times }
% + 2a\mu^2 \,  \frac{p_\nu}{p^2}\, \frac{\epsilon^{\gamma\delta} k_\gamma l_\delta}{k^2 l^2}\,
%  \sin \frac{r\wedge p}{2}\, \sin \frac{l\wedge k}{2}\,  \Big)  
%  \nonumber \\[8pt]
% &&\phantom{{\mu^2}{(2\pi)^6} }
=\int dp \,dk\,dl \, \, \delta (p+k+l-r) \, \cK(s,p,k,l)  \nonumber \\[8pt]
&&\phantom{{\mu^2}{(2\pi)^6} }
\times\Big( 4 \,  \frac{\tilde p_\nu}{p^2} \,(
 \cos \frac{r\wedge p}{2}\, \cos\frac{l\wedge k}{2} -\sin \frac{r\wedge p}{2}\, \sin \frac{l\wedge k}{2} )
- \frac{p_\nu}{p^2} \, \sin \frac{r\wedge p}{2}\,
\cos \frac{l\wedge k}{2} 
\nonumber\\[8pt]
&&\phantom{{\mu^2}{(2\pi)^6}\times }
- a \, \big( 4 \,  \frac{\tilde p_\nu}{p^2} \,
 \cos \frac{k\wedge p}{2}\, \cos\frac{l\wedge r}{2}
- 2\mu^2 \,  \frac{p_\nu}{p^2}\, \frac{ k \cdot \tilde l}{k^2 l^2}\,
 \sin \frac{r\wedge p}{2}\, \sin \frac{l\wedge k}{2}\, \big) \Big)  
  \\[16pt]
 &&\bbKK(r,s) 
%=\int dp \,dq\, \, \frac{2}{q^2} \, K(-r,-p)K(-s,p)  \nonumber\\[8pt]
% &&\phantom{{\mu^2}{(2\pi)^6}}
% \times\Big( 2 ( \, \sin^2 \frac{q\wedge p}{2}
% + \epsilon^2 \cos^2 \frac{q\wedge p}{2}) + a\mu^2 \, \frac{1}{p^2}\,
% \sin^2 \frac{q\wedge p}{2} \Big) 
% % \nonumber\\[8pt]
% &&\phantom{{\mu^2}{(2\pi)^6}\times }
=\int dp \,dq\, \, \frac{2}{q^2} \, K(r,-p)K(s,p)   \, \Big( 2  - a \, ( 
\cos^2 \frac{q\wedge p}{2} - \frac{\mu^2 }{p^2}
\sin^2 \frac{q\wedge p}{2} ) \Big) \,
\\[16pt]
 &&\mathbb{Q}(r) =  \frac{1}{r^2}\,
\int  \frac{dq}{q^2}\,  \sin^2 \frac{q\wedge r}{2}
\\[16pt]
 && \cJ(s) =\int dk\,dl\,\delta(k+l-s) \, \cos \frac{k\wedge l}{2}\,  K(k,l).
\end{eqnarray}
The propagator corrections are then given by
\begin{eqnarray}
&&\hskip-0.9cm       
 P_{\phi(r)\phi(s)} = \frac {\mu^2}{(2\pi)^6}\, \bbN(r,s) +\frac{2}{(2\pi)^4 a} \, \bbKK(r,s) 
 \\[20pt]
 &&\hskip-0.9cm P_{\phi(r)A_\mu(s)}
%  = -\frac {i\mu^3}{(2\pi)^6}\frac{\tilde s_\mu}{s^2}
% \, \bbN(r,s)
% -\frac {2 i\mu}{(2\pi)^4 a}\frac{\tilde s_\mu}{s^2}\,
% \bbKK(r,s)     \\[8pt]
% &&\phantom{ \ \ \ \ \ }
% -\frac {i\mu}{(2\pi)^4 a}\, \frac{1}{s^2}\, \bbK_\mu(r, s)
% +\frac {4i\mu}{(2\pi)^2 a}\frac{\tilde s_\mu}{s^2}\,  K(r,s)\, \bbQ(s)    \nonumber\\[12pt]
% &&\phantom{ \ \ \ \ \ }
=-i\mu \, \frac{\tilde s_\mu}{s^2}\, P_{\phi(r)\phi(s)}
-\frac {i\mu}{(2\pi)^4 a}\, \frac{1}{s^2}\, \bbK_\mu(r, s)
+\frac {4i\mu}{(2\pi)^2 a}\frac{\tilde s_\mu}{s^2}\,  K(r,s)\, \bbQ(s)
  , \\[20pt]
&&\hskip-0.9cm P_{A_\nu(r)A_\mu(s)} = 
%- \frac {\mu^4}{(2\pi)^6}
% \frac{\tilde s_\mu}{s^2} \frac{\tilde r_\nu}{r^2}\, \bbN(r,s)
% - \frac{ 2 \mu^2}{(2\pi)^4 a}\, \frac{\tilde s_\mu}{s^2}\,
% \frac{\tilde r_\nu}{r^2}\, \bbKK(r,s)             \\[8pt]
% &&\phantom{ \ \ \ \ \ }
% - \frac {\mu^2}{(2\pi)^4 a}\,\frac{\tilde s_\mu}{s^2}\,
% \frac{1}{r^2}\, \bbK_\nu (s,r)
% - \frac {\mu^2}{(2\pi)^4 a}\,\frac{\tilde r_\nu}{r^2}\,
% \frac{1}{s^2}\, \bbK_\mu(r,s)
%  \nonumber\\[8pt]
% %
% &&\phantom{ \ \ \ \ \ }
% +\frac {4}{a^2 } \,
% \frac{\delta_{\nu\mu} \delta(r+s)}{s^2}\, \bbQ(r)
% + \frac{4\mu^2}{(2\pi)^2 a}\,
% \frac{\tilde s_\mu}{s^2}\,
% \frac{\tilde r_\nu}{r^2}\, K(r,s) \,\big(  \bbQ(r)+\bbQ(s) \big)
%  \nonumber\\[8pt]
% % K
% &&\phantom{ \ \ \ \ \ }
% + \frac {1}{(2\pi)^2 a^2}\, \,
% \frac{1}{r^2 s^2}
% \int \,dk\,dl \, \delta (k+l-r-s) \, K(k,l)  \nonumber \\[8pt]
% &&\phantom{ \ \ \ \ \  }
%  \times \Big( 
% -4 \delta_{\mu\nu}
%  \,(\, \sin \frac{s\wedge r}{2}\, \sin \frac{l\wedge k}{2} 
% + \cos \frac{s\wedge r}{2}\, \cos\frac{l\wedge k}{2} )
% - \epsilon_{\nu\mu}
%  \,\sin \frac{s\wedge r}{2}\,
% \cos \frac{l\wedge k}{2} \nonumber \\[8pt]
% &&\phantom{ \ \ \ \ \ \ }
% + a \big( 4 \delta_{\mu\nu}  \cos \frac{k\wedge r}{2}\, \cos\frac{l\wedge s}{2}
% + 4\mu^2 \frac{k_\nu l_\mu}{k^2 l^2} \, \sin \frac{k\wedge r}{2} \,\sin \frac{l\wedge s}{2}
% %  \nonumber \\[8pt]
% % &&\phantom{ \frac{\mu^2}{(2\pi)^6}\times \times \ \ \  }
% +2\mu^2 \epsilon_{\nu\mu}\,  \frac{k \cdot\tilde l}{k^2 l^2} \,
% \sin \, \frac{s\wedge r}{2} \, \sin \frac{l\wedge k}{2}\,
% \big)  \Big) 
% \nonumber \\[12pt]
% &&\phantom{ \ \ \ \ \  }
 \mu^2 \, \frac{\tilde s_\mu}{s^2}\, \frac{\tilde r_\nu}{r^2}\,P_{\phi(r)\phi(s)}
-i\mu \, \frac{\tilde s_\mu}{s^2}\, P_{\phi(s)A_\nu(r)} 
-i\mu \, \frac{\tilde r_\nu}{r^2}\,  P_{\phi(r)A_\mu(s)} +\frac {4}{a^2 } \,
\frac{\delta_{\nu\mu} \delta(r+s)}{s^2}\, \bbQ(r)
\\[8pt] 
&&\hskip-0.9cm\phantom{ \ \ \ }
+ \frac {1}{(2\pi)^2 a^2}\, \,
\frac{1}{r^2 s^2}
\int \,dk\,dl \, \delta (k+l-r-s) \, K(k,l)  \nonumber \\[8pt]
&&\hskip-0.9cm\phantom{ \ \ \   }
 \times \Big( 
-4 \delta_{\mu\nu}
 \,(\, \sin \frac{s\wedge r}{2}\, \sin \frac{l\wedge k}{2} 
+ \cos \frac{s\wedge r}{2}\, \cos\frac{l\wedge k}{2} )
- \epsilon_{\nu\mu}
 \,\sin \frac{s\wedge r}{2}\,
\cos \frac{l\wedge k}{2} \nonumber \\[8pt]
&&\hskip-0.9cm\phantom{ \ \ \ \  }
+ a \big( 4 \delta_{\mu\nu}  \cos \frac{k\wedge r}{2}\, \cos\frac{l\wedge s}{2}
+ 4\mu^2 \frac{k_\nu l_\mu}{k^2 l^2} \, \sin \frac{k\wedge r}{2} \,\sin \frac{l\wedge s}{2}
%  \nonumber \\[8pt]
% &&\phantom{ \frac{\mu^2}{(2\pi)^6}\times \times \ \ \  }
+2\mu^2 \epsilon_{\nu\mu}\,  \frac{k \cdot\tilde l}{k^2 l^2} \,
\sin \, \frac{s\wedge r}{2} \, \sin \frac{l\wedge k}{2}\,
\big)  \Big)  .
\nonumber 
\end{eqnarray}

As before,  to obtain  divergent parts  we amputate the external propagator 
legs by multiplying  from the left and from the right by the inverse propagator; the 
amputated 2-point function, a 2$\times $2 matrix $\Pi(r,s)$, is
\begin{equation}
 \Pi(r,s)
= \frac{1}{(2\pi)^4} \int dp\, dq \, G^{-1}(r, -p)   \label{pi}
P(p,q)\,G^{-1}(-q,s) .
\end{equation}
Again we  first calculate the $1/a$- and $a$-constant divergent parts of (\ref{pi})
which we denote by  $\Pi\vert\, $.  We obtain
\begin{eqnarray}
 &&\hskip-1.6cm
\Pi_{\rho\sigma}(\rho r,\sigma s)\vert =4 \delta_{\rho\sigma}\delta(r+s) s^2 \bbQ (s) 
-\frac{1}{(2\pi)^2}\, (4\delta_{\rho\sigma} \cos \frac{r\wedge s}{2} 
-\epsilon_{\rho\sigma} \sin \frac{r\wedge s}{2})\, \cJ(r+s)  \\[8pt]
&&\hskip-1.6cm
\Pi_{\rho} (\rho r,s)\vert=
\frac{2i\mu}{(2\pi)^2}\, \int  dk\,dl\,\delta(k+l-r-s) \, \cos \frac{s\wedge l}{2}\,  K(k,l)
\big(  \frac{4\tilde k_\rho }{k^2}\,     \cos \frac{r\wedge k}{2}  
-\frac{k_\rho }{k^2}\,     \sin \frac{r\wedge k}{2}  
 \big)             \\[8pt]
&&\hskip-1.6cm  \Pi(r,s)\vert=
\frac{8}{a}\, \delta(r+s)\int \frac{dl}{l^2} 
-4\, \delta(r+s)\int \frac{dl}{l^2}\cos^2\frac{r\wedge l}{2}   \\[8pt]
&& \hskip-1.6cm   \phantom{\tau_{\phi(r)\phi(s)}}
+\frac{\mu^2}{(2\pi)^2}\, \cos\frac{r\wedge s}{2}\,\int dk\, dl\, \delta(k+l-r-s)
\Big( 4\frac{k\cdot l}{k^2 l^2} \cos \frac{k\wedge l}{2} 
-\frac{k \cdot\tilde l}{k^2 l^2} 
\sin\frac{k\wedge l}{2}\Big) K(k,l) .  \nonumber
\end{eqnarray}
These integrals can be naturally rewritten using the `short variable'
$u=r+s$, the difference between the incoming and the
outgoing momentum, and the  `long variable' $v=r-s$ which is done
 in Appendix II.
 
The $a$-linear part $\Delta\Pi$ is given by
\begin{eqnarray}
 &&\hskip-1cm
\Delta\Pi_{\rho\sigma}=\frac{a}{(2\pi)^2}\int dk\, dl\, \delta(k+l-r-s) \, K(k,l)
\\[8pt]
&&\hskip-1cm \phantom{\Delta\Pi_{\rho\sigma}}
\times\Big( 4\delta_{\rho\sigma}\cos\frac{k\wedge r}{2}\cos\frac{l\wedge s}{2}
+ 4\mu^2\frac{k_\rho l_\sigma}{k^2 l^2}\, \sin\frac{k\wedge r}{2}\sin\frac{l\wedge s}{2}
+ 2\mu^2\epsilon_{\rho\sigma}\frac{k\cdot\tilde l}{k^2 l^2}\,\sin\frac{s\wedge r}{2}\sin\frac{l\wedge k}{2}
\Big)\nonumber \\[8pt]
&&\hskip-1cm
\Delta\Pi_\sigma = -\frac{4ia\mu}{(2\pi)^2}\int dk\, dl \,\delta(k+l-r-s)\, K(k,l)\,
\frac{\tilde k_\sigma}{k^2}\, \cos\frac{k\wedge r}{2}\cos\frac{l\wedge s}{2}
\\[8pt]
&&\hskip-1cm \phantom{\Delta\Pi_\sigma =}
-\frac{4ia\mu}{(2\pi)^2}\int dk\, dl \,\delta(k+l-r-s)\, K(k,l)\,
\frac{\tilde k_\sigma}{k^2}\, \cos\frac{k\wedge l}{2}\cos\frac{r\wedge s}{2}
\nonumber \\[8pt]
&&\hskip-1cm
\Delta\Pi = \frac{4\mu^2 a}{(2\pi)^2}\int dk\, dl \, \delta(k+l-r-s)\, K(k,l)\,
\frac{k\cdot l}{k^2 l^2}\,\cos\frac{k\wedge r}{2}\cos\frac{l\wedge s}{2}  .
\end{eqnarray}

The one-loop contributions to the propagators are now rather long integrals.
Therefore the calculation and analysis of the corresponding divergences is placed in
Appendix II. It is interesting to mention however that many terms cancel in the course of calculation, and 
in  very encouraging way. At the end we obtain only three
 divergent contributions to the 2-point functions at one loop; the corresponding counterterms are
\begin{equation}
 \int d^2x \, A_\mu\star  A^\mu, \qquad \int d^2x\, 
 \phi\star \phi \qquad {\rm and}\qquad \int d^2x\,\{ \tilde x^\mu\ds A_\mu \}\star\phi .
                                                                                  \label{propag}
\end{equation}
Again the forefactors are proportional to
 integrals which diverge logarithmically.

\initiate
\section{ Conclusions and outlook}

As we said at the beginning, our paper is devoted to an analysis
of renormalizability properties of the BGM model \cite{Buric:2010xs}
which is a gauge analogon of the GW model.

The classical model was constucted in  \cite{Buric:2010xs} and some of its 
propeties were explored there: the equations of motion, the vacuum solutions, the
BRST symmetry. Here we study perturbative quantization of the model.
The model is apparently quite complicated:
it contains a scalar field and a gauge field mixed already at the level of
propagators; the interaction is described by ten vertices. Therefore the 
renormalizability analysis has not been completed yet, though a considerable amount
of work has been done here.

Let us resume it shortly. First, the model was represented on the 
two-dimensional Moyal space and then the Feynman rules
of the theory were derived. As it was impossible to diagonalize the 
kinetic term, we treated the fields as a multiplet,
as in supersymmetry. The propagator became a $2 \times 2$ matrix
 containing  the Mehler kernels in all matrix elements, which means
that the background curvature influences propagation of all fields.
 We then calculated the one-loop quantum corrections to the
tadpoles and to the propagators, leaving the vertex corrections for the subsequent work.
The quantum corrections which we obtained are divergent: all logarithmically.
Notably the tadpole terms do not vanish. This property can be related to the
non-conservation of the momentum; it appears in similar models,
 \cite{deGoursac:2007qi,Grosse:2007dm,Blaschke:2007vc,
deGoursac:2008rb,Blaschke:2009aw}. Expressed as counterterms, the 1-point function
divergences are $\, \int \phi \,  $  and $\, \int \tilde x_\mu\star A^\mu \, $. Obviously these
terms are not present in the initial action (\ref{modelM}). The 2-point functions also have 
divergent corrections, 
$\, \int \phi\star\phi \,  $, $\, \int A_\mu\star A^\mu  \, $ and 
 $\, \int \{ \tilde x_\mu\ds A^\mu\}\star \phi  \, $.

There are two ways to understand counterterm  (\ref{tadpole}) and (\ref{propag}) in our model. 
One possibility is to interpret these terms as  indication that the trivial
vacuum $\phi =0$, $A_\mu =0$ we started with is unstable under quantization,
and that the  quantum vacuum is of the form
\begin{equation}
 \phi = \alpha,\qquad A_\mu =\beta\tilde x_\mu,                  \label{*}
\end{equation}
which the second classical vacuum of our theory has.
Expansion around (\ref{*}) obviously gives all terms which we obtained
as divergences, and some additional ones. The second possibility is that
all counterterms add up to  Chern-Simons action (\ref{modelCS}): this would mean
 that $\cS_{CS}$ should be included in the classical action.

To complete our analysis and come to conclusive results we have to perform a
couple more steps. First, we need to calculate corrections to the vertices:
this will help us to decide whether the origin of  divergences
is a shift of the vacuum or the Chern-Simons term (or both). 
 Also, to obtain and compare the coefficients in the counterterms one
should find a systematic way to quantify divergences in the parameter 
integrals. At present, we were able only to analyze the type of divergences, as
 when there are two parameter integrals the expressions
are quite complicated and as a rule, impossible to solve exactly
 in terms of special functions.
One should also do the power counting and estimate  higher-order contributions.
 And finally, it is to be expected that
 for particular values of  parameters $a$, $\mu$ and $c$ our model has specific
renormalizability properties. All these points we plan to address in our
future work.

\vskip1cm \noindent
{\bf Acknowledgement}
\hskip0.3cm
The work of M.~B., M.~D. and V.~R. is supported by project
171031 of the Serbian Ministry of Education and Science.

\initiate
\section{Appendix I}

\noindent
{\bf i) Fourier transformation conventions}
\begin{eqnarray}
 &&  \nonumber
 f(x) = \frac{1 }{(2\pi)^2} \int d^2p\, \tilde f(p) e^{-ipx }\,\\[4pt]
&& \nonumber
f(x,y) =  \frac{1}{(2\pi)^4} \int  d^2p \,d^2q\,  \tilde f(p,q) e^{-ipx - iqy}\,.
\nonumber
\end{eqnarray}
With this convention in the  $n$-point functions all
momenta are incoming. Thus for example the Fourier transformation
of the identity operator is
\begin{eqnarray}
&& I(x,y)= \delta(x-y) =  \frac{1}{(2\pi)^4} \int d^2 p \, d^2q \, \tilde I(p,q) e^{-ipx - iqy}
  \nonumber\\[4pt]
&&\nonumber
 \tilde I(p,q) = (2\pi)^2 \delta(p+q)
\\[8pt] \nonumber
&&
 \widetilde{FG}(p,q) = \frac{1}{(2\pi)^2} \int d^2 r\, \tilde F(p,r)\tilde G(-r,q)
\end{eqnarray}
A useful formula is
\begin{equation}
 \int d^2p \, K^{-1}(r,-p)\, \cK(p,q,k,l) = (2\pi)^4 
\big(\delta(r+q)\, K(k,l) +\delta(r+k) \, K(q,l) + \delta (r+l)\, K(q,k) \big).
\nonumber
\end{equation}

\noindent
{\bf ii) Gaussian integrals}
\begin{eqnarray}
&&
 \frac{1}{p^2} =\int_0^\infty d\beta \, e^{-\beta p^2} \nonumber \\[4pt]
&& \int e^{-ap^2 + bp}\, d^2 p =\frac{\pi }{a } \,  e^{{b^2}/{4a}}
\nonumber \\[4pt]
&& \nonumber
 \int  p_\alpha  e^{-ap^2 + bp}\, d^2 p = \pi \,  \frac{b_\alpha}{2a^2}\,  e^{{b^2}/{4a}}
\\[8pt]
&& \nonumber
\int  \frac {p_\alpha }{p^2} \, e^{-ap^2 + bp} \, d^2 p  = 2\pi\, \frac{ b_\alpha}{b^2} \,  e^{{b^2}/{4a}} 
\\[8pt]
&&\nonumber
\int   \frac {p_\alpha p_\beta }{p^2} \, e^{-ap^2 + bp}\, d^2 p = \,\frac {2\pi}{b^2}\, \Big( 
\frac{\delta_{\alpha\beta}b^2 -2 b_\alpha b_\beta}{b^2} +\frac{b_\alpha b_\beta}{2a }
 \Big) \,  e^{{b^2}/{4a}} 
\end{eqnarray}

\noindent
{\bf iii) Exponential integrals}
\begin{eqnarray}
 && E_\nu(z) = \int_1^\infty \frac{e^{-zt}}{t^\nu}, \qquad {\rm Re} z>0
\nonumber \\[6pt]
&&  E_\nu(z) \sim \frac{e^{-z}}{z}\big( 1 + O(\frac 1z)\big) , \qquad \vert z\vert \to\infty
\nonumber \\[6pt]
&&  E_n(z) =\frac{(-z)^{n-1}}{(n-1)!} \big( \psi(n) -\log z\big) -\sum_{k=0}^\infty \frac{(-z)^k}{(k-n+1)k!}
\nonumber
\end{eqnarray}

\initiate
\section{Appendix II}

Let us discuss some of the integrals which appear in the one-loop propagator
divergences and obtain the corresponding counterterms.  
The  $\,\Pi_{\rho\sigma}\vert\,$ is the simplest as it contains divergent integral which
appeared before,
\begin{eqnarray}
&& \cJ(u) = \int dk\, dl\, \delta(k+l-u)\cos\, \frac{k\wedge l}{2} \, K(k,l) 
= -\frac{\pi^2}{2\mu^2}\int_1^\infty d\xi\, \frac{\xi -1}{\xi +1}\, e^{-\frac{u^2}{4\mu^2}\xi}.
\nonumber 
% &&\ \ 
% = -\frac{\pi^2}{2\mu^2} \, e^{\frac{u^2}{4\mu^2}}
% \Big( E_0(\frac{u^2}{2\mu^2}) -E_1(\frac{u^2}{2\mu^2})\Big)  
% =-\frac{\pi^2}{2\mu^2} \, e^{\frac{u^2}{4\mu^2}}
% \Big(\frac{2\mu^2}{u^2}\,  e^{-\frac{u^2}{2\mu^2}}+\gamma+\log \frac{u^2}{2\mu^2}\dots
%  \Big)     .   \nonumber
\end{eqnarray}
The second component,  $\,\Pi\vert\,$, or more precisely its part
 containing the Mehler kernel is given by
\begin{eqnarray}
&&
\frac{\mu^2}{(2\pi)^2}\, \cos\frac{r\wedge s}{2}\,\int dk\, dl\, \delta(k+l-r-s)\,
\Big( 4\frac{k\cdot l}{k^2 l^2} \cos \frac{k\wedge l}{2} 
-\frac{k\cdot \tilde l}{k^2 l^2} 
\sin\frac{k\wedge l}{2}\Big) K(k,l) \nonumber   \\[8pt]
&& \ \  =
-\frac{1}{8\mu^2}\cos\, \frac{r\wedge s}{2}\,
 \iint_1^\infty d\xi\, d\eta\, \frac{\xi -1}{\xi+1} \,
 \frac{4\eta -2 + 2\xi^2 -\xi\eta}{\xi\eta\big( (2\eta -1)^2 -\xi^2\big)}\,
e^{-\frac{u^2}{8\mu^2} \, \big(\xi +\frac1\xi +\frac1\eta (\xi -\frac1\xi)\big)} ,
\nonumber
\end{eqnarray}
 and finally, in the $\, \Pi_\rho\vert\,$ propagator partn the
integration gives using $\ 2 s\wedge r =- u\wedge v$,
\begin{eqnarray}
&&\frac{2i\mu}{(2\pi)^2}\, \int  dk\,dl\,\delta(k+l-r-s) \, \cos \frac{s\wedge l}{2}\,  K(k,l)
\big(  \frac{4\tilde k_\rho }{k^2}\,     \cos \frac{r\wedge k}{2}  
-\frac{k_\rho }{k^2}\,     \sin \frac{r\wedge k}{2}  
 \big)      
\nonumber\\[8pt]
&&\ \ 
 = \frac{i(2\pi)^4}{4\mu}\, \frac{1}{u^2}\int_1^\infty \frac{d\xi}{(\xi+1)^2}\,
e^{-\frac{u^2}{4\mu^2}\,\xi}\Big( (4-\xi)\tilde u_\rho \cos\frac{u\wedge v}{4}\, +
(4\xi -1) u_\rho \sin\frac{u\wedge v}{4}\Big) \nonumber \\[8pt]
&& \ \  
-\frac{i(2\pi)^4}{4\mu}\int_1^\infty d\xi\,  \frac{\xi-1}{\xi+1}\,
e^{-\frac{1}{8\mu^2}\,(u^2 +v^2 )\xi} \, \frac{1}{(u-\xi v)^2}\, \frac{1}{(u-\xi v)^2}
\nonumber\\[8pt]
&&\ \ \ \
\times \Big(\cos\frac{ u\wedge v}{2} \, \big( (4\tilde u_\rho -\xi \tilde v_\rho)(u^2 -\xi^2 v^2) 
+2(u_\rho -4\xi v_\rho) u\cdot\xi \tilde v\big) \nonumber\\[8pt]
&&\ \  \ \  \ \ 
+ \sin \frac{u\wedge v}{2}\, \big( (u_\rho -4\xi v_\rho)(u^2 -\xi^2 v^2)  -
2 (4\tilde u_\rho -\xi v_\rho)u\cdot\xi \tilde v\big) \Big) . \nonumber
\end{eqnarray}

Let us analyze the corresponding counterterms.
We start with the simplest, $\Pi_{\rho\sigma}\vert  $: as before, 
we multiply it by the external fields and integrate.
The first summand of the $\Pi_{\rho\sigma}\vert$ has a $\delta$-function;
of course there is no overall
translation invariance, and the momentum conservation is broken by the second part
of $\Pi_{\rho\sigma}\vert$, in which the Mehler kernel plays a role of a smeared
  $\delta(r+s)$. Denoting
\begin{equation}
 \iint dr\,ds\, A^\rho(r)\Pi_{\rho\sigma}(-r,-s) A^\sigma(s)
=  \iint dr\,du\, A^\rho(r) A^\sigma(-r+u) \Pi_{\rho\sigma}(-r,r-u) \equiv
(\oldstylenums{1})+(\oldstylenums{2}) , \nonumber
\end{equation}
where $u= r+s$, we have
\begin{eqnarray}
 &&(\oldstylenums{1})= 2 \iint d^2r \,d^2u\, A^\rho(r) A_\rho(-r+u)\, \delta(u)
\int \frac{d^2q}{q^2}(1-\cos (q\wedge r))          \nonumber
\\[8pt] &&\quad
=4\pi\int d^2r \, A^\rho(r) A_\rho(-r) \int_0^\infty  \frac{dq}{q}
-2\int d^2r \, A^\rho(r) A_\rho(-r)\int d^2q\,\int_0^\infty d\beta\,e^{-\beta q^2}\cos (q\wedge r)
\nonumber \\[8pt] &&\quad
=2\pi\, \Big( 2 \int_0^\infty  \frac{dq}{q}-\Gamma(0)\Big) \, \int d^2r \, A^\rho(r) A_\rho(-r).
\nonumber
\end{eqnarray}
The second part is equal to
\begin{eqnarray}
&& \hskip-1.5cm (\oldstylenums{2})=-\frac{1}{(2\pi)^2}\,\iint d^2r \,d^2u\, A^\rho(r) A^\sigma(-r+u)
(4\delta_{\rho\sigma} \cos \frac{r\wedge u}{2} 
-\epsilon_{\rho\sigma} \sin \frac{r\wedge u}{2})\,\cJ(u)     \nonumber  \\[8pt]
         &&\hskip-1.5cm\quad
= \frac{1}{8\mu^2} \iint d^2r \,d^2u\, A^\rho(r) \big( A^\sigma(-r) +\p^\alpha A^\sigma(-r) u_\alpha
+\dots\big) 
\big(4\delta_{\rho\sigma} + \epsilon_{\rho\sigma} \frac{\tilde r u}{2\mu^2}+\dots\big)
\nonumber \\[8pt]
&&\hskip-1.5cm\phantom{(\oldstylenums{2})=-\frac{1}{(2\pi)^2}  } \times
\, e^{\frac{u^2}{4\mu^2}}
\big(\frac{2\mu^2}{u^2}\,  e^{-\frac{u^2}{2\mu^2}}+\gamma+\log \frac{u^2}{2\mu^2}+\dots
 \big)      .  \nonumber
\end{eqnarray}
The leading divergence in the last expression is
$ \ 2\pi\,  \int_0^\infty  \frac{dq}{q} \, \int d^2r \, A^\rho(r) A_\rho(-r)\, $,        
 the remaining terms are all convergent. Therefore we conclude that
the divergent part of $\Pi_{\rho\sigma}$ gives a counterterm proportional to
\begin{equation}
\frac{1}{(2\pi)^2} \int d^2r \, A^\rho(r) A_\rho(-r) = \int d^2x \, A^\rho(x) A_\rho(x) .          \label{a^2}
\end{equation}
The divergence is logarithmic.

The second counterterm can also be divided into two:
\begin{equation}
 \iint dr\,ds\, \phi(r)\,\Pi(-r,-s) \, \phi(s)
=  \iint dr\,du\, \phi(r)\, \phi(-r+u)\, \Pi(-r,r-u) \equiv
(\oldstylenums{3})+(\oldstylenums{4}),  \nonumber
\end{equation}
where
\begin{eqnarray}
 &&\hskip-1cm (\oldstylenums{3})=  \iint d^2r \,d^2u\, \phi(r) \, \phi(-r+u)\, \delta(u)
\Big( (\frac 8a -2) \int \frac{d^2q}{q^2}-2 \int \frac{d^2q}{q^2}\, \cos (q\wedge r)\Big) 
\nonumber    \\[8pt] &&\hskip-1cm \quad
=2\pi\, (\frac 8a -2)\,\int_0^\infty \frac{dq}{q}\,\int d^2r \, \phi(r) \, \phi(-r) .
\nonumber
\end{eqnarray}
The other part is
\begin{eqnarray}
&&(\oldstylenums{4}) =-\frac{1}{8(2\pi)^2}\iint d^2r\,d^2u\,
\phi(r)\, \phi(-r+u)\,\cos\frac{r\wedge u}{2}                   \nonumber
 \\[8pt]
&&\qquad \qquad
\times\iint_1^\infty d\xi\, d\eta\,\frac{\xi-1}{\xi+1}
\, \frac{4\eta -2+2\xi^2-\xi\eta}{\xi\eta \big( (2\eta -1)^2-\xi^2\Big)}\,
e^{-\frac{u^2}{8\mu^2}\big( \xi +\frac 1\xi +\frac 1\eta (\xi -\frac 1\xi ) \big)}
\nonumber \\[8pt]
&&\qquad =-\frac{1}{8(2\pi)^2}\iint d^2r\,d^2u\,
\phi(r)\, \big( \phi(-r) +\p_\alpha\phi(-r)\, u^\alpha +\dots \big)\,  
\big(1 +\frac{1}{4\mu^2} \tilde r u +\dots \big)
\nonumber \\[8pt]
&&\qquad \qquad
\times\iint_1^\infty d\xi\, d\eta\,\frac{\xi-1}{\xi+1}\, \frac{1}{\xi\eta}\,
 \frac{4\eta -2+2\xi^2-\xi\eta}{ (2\eta -1)^2-\xi^2}\,
e^{-\frac{u^2}{8\mu^2}\big( \xi +\frac 1\xi +\frac 1\eta (\xi -\frac 1\xi ) \big)}   \, .
\nonumber
\end{eqnarray}
The leading potentially divergent term after the change of variables
 $2\eta -1 =\zeta\,$ becomes
\begin{eqnarray}
 &&\hskip-20pt
-\frac{1}{8(2\pi)^2}\int d^2r\,
\phi(r)\,  \phi(-r) \int 2\pi u\,  du
\iint_1^\infty  d\xi\, d\eta\,\frac{\xi-1}{\xi+1}
\, \frac{4\eta -2+2\xi^2-\xi\eta}{\xi\eta \big( (2\eta -1)^2-\xi^2\big)}\,
e^{-\frac{u^2}{8\mu^2}\big( \xi +\frac 1\xi +\frac 1\eta (\xi -\frac 1\xi ) \big)}   
\nonumber\\[8pt]
&&\hskip-18pt
= -\frac{\mu^2}{8\pi}\int d^2r\,
\phi(r)\,  \phi(-r) 
\iint_1^\infty d\xi\, d\zeta\, \, \frac{\xi-1}{\xi+1}\,
\, \frac{4\zeta +4\xi^2-\xi (\zeta+1)}{ \big( \zeta^2-\xi^2\big) \big((\zeta +1)(\xi^2+1)+2\xi^2 -2  \big)}\,.
\nonumber
\end{eqnarray}
and it is finite. Therefore the divergence comes only from 
$ (\oldstylenums{3}) $ and  it is proportional to
\begin{equation}
\frac{1}{ (2\pi)^2} \int d^2 r \, \phi(r) \, \phi(-r) = \int d^2x \, \phi(x) \, \phi(x)   .         \label{fi^2}
\end{equation}
Again it is logarithmic.

By similar reasoning we find that there are
no divergences in  $\Pi_\rho\vert$ terms; however, in the $a$-linear part of the 2-point function
a divergent term appears. We denote
\begin{equation}
\iint dr\, ds A^\rho(r)\Pi_\rho(-r, -s)\phi(s) = (\oldstylenums{5}) +
(\oldstylenums{6}).   \nonumber
\end{equation}
Part $(\oldstylenums{5})$ is finite, while for part $(\oldstylenums{6})$ we obtain
\begin{eqnarray}
(\oldstylenums{6}) &=&
-\frac{4i\mu a}{(2\pi)^2}\iint dr\, ds\, A^\rho(r)\phi(s)\iint dk\, dl\,
\delta(k+l-r-s)\, K(k,l)\, \frac{\tilde{k}_\rho}{k^2} \, 
\cos\frac{k\wedge l}{2}\cos\frac{r\wedge s}{2} \nonumber\\
&=& \frac{ia}{16\mu}\iint dr\, ds A^\rho(r)\phi(s) \cos\frac{r\wedge s}{2}\,
\frac{\tilde{u}_\rho}{u^2}\int_1^\infty d\xi (\xi -1) 
e^{-\frac{u^2}{8\mu^2}(1+\epsilon^2)\xi} \nonumber\\
&&\times\Big(\frac{1}{(1-\epsilon^2)(1+\xi)} + \frac{\epsilon}{2(\epsilon -1)(1+\epsilon\xi)}
+ \frac{\epsilon}{2(\epsilon +1)(1-\epsilon\xi)}\Big)  . \nonumber
\end{eqnarray}
After the parameter integrations the only divergent contribution is of the form 
\begin{equation}
(\oldstylenums{6}) =
-\frac{i\mu a\epsilon^2}{2}\iint dr\, du\, A^\rho(r)\phi(-r+u)\cos\frac{r\wedge u}{2}
\,\frac{\tilde{u}_\rho}{u^4} \,e^{-\frac{1+\epsilon^2}{8\mu^2}u^2} .\nonumber
\end{equation}
As usual, we expand around $u=0$ and obtain
\begin{equation}
(\oldstylenums{6}) =
-\frac{i\mu a\epsilon^2}{2}\iint dr\, ds\, A^\rho(r)\Big( \phi (-r) + u^\alpha 
\partial_\alpha \phi(-r) + \dots \Big)\, \frac{\tilde{u}_\rho}{u^4} .\nonumber
\end{equation}
Using the polar coordinates we see that the first term gives vanishing contribution while the 
second term is
\begin{equation}
- \frac{i}{(2\pi)^2} \int d^2r\, \epsilon^{\rho\alpha}A_\rho(r) \partial_\alpha \phi(-r) = 
\int d^2x\,\tilde x^\rho A_\rho\phi .\label{propNew}
\end{equation}
The divergence is again logaritmic.

Infinities  appear also in $\Delta\Pi_{\rho\sigma}$ and $\Delta\Pi$ but they are  of the same
forms (\ref{a^2}) and (\ref{fi^2}).

% 
% It is interesting that if we introduce in the second integral $w=u-\xi v$, $z =u+\xi v$, 
% the exponential factor obtains the form of the Mehler kernel in $w,\ z$: as if there is
% some scaling.  The polynomials can be rewritten too, but then what I don't know
% \begin{eqnarray}
%  &&I_2 =  -\frac{i(2\pi)^4}{8\mu}\int_1^\infty d\xi\, \frac{\xi-1}{\xi+1}\,
% e^{-\frac{1}{32\mu^2}\,\big((w+z)^2\xi +(w-z)^2\frac1\xi\big)} \, \frac{1}{w^2 z^2}
% \nonumber\\[8pt]
% &&\phantom{\quad}
% \times \Big(\cos \frac{w\wedge z}{4\xi} \big( (5\tilde w_\rho +3 \tilde z_\rho) w\cdot z
% +(5 w_\rho -3 z_\rho) w\cdot\tilde z\big) \nonumber\\[8pt]
% &&\phantom{\quad \quad}
% +\sin \frac{w\wedge z}{4\xi} \big( (5w_\rho -3 z_\rho)w\cdot z -
%  (5\tilde w_\rho +3\tilde z_\rho)  w\cdot\tilde z\big) \Big) .
% \end{eqnarray}
% 
% % 
% 
% \bibliographystyle{../latex-styles/utphys}
% \bibliography{../tuw}

\end{document}